\documentclass[twocolumn]{aastex63}
\pdfoutput=1 
\usepackage{amsmath,amstext}
\usepackage[T1]{fontenc}
\usepackage{apjfonts} 
\usepackage[figure,figure*]{hypcap}
\usepackage{upgreek}
\usepackage{float}
\usepackage{rotating}
\usepackage{boldline}
\usepackage{listings}
\usepackage{hyperref}
\usepackage{xcolor}

\usepackage{xspace}
\newcommand{\Teff}{$T_\mathrm{eff}$\xspace}
\newcommand{\Lbol}{$L_\mathrm{bol}$\xspace}
\usepackage{booktabs}

\graphicspath{{./}{figures/}}

\received{August 4, 2021}
\revised{September 21, 2021}
\accepted{\today}
\submitjournal{ApJ}

\shorttitle{Retrieval of SDSS J1256$-$0224}
\shortauthors{Gonzales et al.}

\begin{document}

\title{The first retrieval of a substellar subdwarf: A cloud-free SDSS J125637.13$-$022452.4}

\correspondingauthor{Eileen Gonzales}
\email{ecg224@cornell.edu}

\author[0000-0003-4636-6676]{Eileen C. Gonzales}
\altaffiliation{51 Pegasi b Fellow}
\affiliation{Department of Astronomy and Carl Sagan Institute, Cornell University, 122 Sciences Drive, Ithaca, NY 14853, USA}
\affiliation{Department of Astrophysics, American Museum of Natural History, New York, NY 10024, USA}
\affiliation{The Graduate Center, City University of New York, New York, NY 10016, USA}
\affiliation{Centre for Astrophysics Research, School of Physics, Astronomy and Mathematics, University of Hertfordshire, Hatfield AL10 9AB}
\affiliation{Department of Physics and Astronomy, Hunter College, City University of New York, New York, NY 10065, USA}

\author[0000-0003-4600-5627]{Ben Burningham}
\affiliation{Centre for Astrophysics Research, School of Physics, Astronomy and Mathematics, University of Hertfordshire, Hatfield AL10 9AB}

\author[0000-0001-6251-0573]{Jacqueline K. Faherty}
\affiliation{Department of Astrophysics, American Museum of Natural History, New York, NY 10024, USA}

\author[0000-0001-6627-6067]{Channon Visscher}
\affiliation{Chemistry \& Planetary Sciences, Dordt University, Sioux Center, IA}
\affiliation{Center for Extrasolar Planetary Systems, Space Science Institute, Boulder, CO}

\author[0000-0002-5251-2943]{Mark Marley}
\affiliation{Department of Planetary Sciences and Lunar and Planetary Laboratory, University of Arizona, Tuscon, AZ}

\author[0000-0003-3444-5908]{Roxana Lupu}
\affiliation{BAER Institute/ NASA Ames Research Center, Moffett Field, CA 94035, USA}

\author{Richard Freedman}
\affiliation{Seti Institute, Mountain View, CA }
\affiliation{NASA Ames Research Center, Moffett Field, CA 94035, USA}

\author[0000-0002-8507-1304]{Nikole K. Lewis}
\affiliation{Department of Astronomy and Carl Sagan Institute, Cornell University, 122 Sciences Drive, Ithaca, NY 14853, USA}

\begin{abstract}
We present the first retrieval analysis of a substellar subdwarf, SDSS J125637.13$-$022452.4 (SDSS J1256$-$0224), using the \textit{Brewster} retrieval code base. We find SDSS J1256$-$0224 is best fit by a cloud-free model with an ion (neutral H, H$^{-}$, and electron) abundance corresponding to [Fe/H]$_{ion}=-1.5$. However, this model is indistinguishable from a cloud-free model with [Fe/H]$_{ion}=-2.0$ and a cloud-free model with [Fe/H]$_{ion}=-1.5$ assuming a subsolar carbon-to-oxygen ratio. We are able to constrain abundances for H$_2$O, FeH, and CrH, with an inability to constrain any carbon bearing species likely due to the low-metallicity of SDSS J1256$-$0224. We also present an updated spectral energy distribution (SED) and semi-empirical fundamental parameters. Our retrieval- and SED-based fundamental parameters agree with the Baraffe low-metallicity evolutionary models. From examining our ``rejected'' models (those with $\Delta$BIC$>45$), we find that we are able to retrieve gas abundances consistent with those of our best fitting model. We find the cloud in these poorer fitting ``cloudy'' models is either pushed to the bottom of the atmosphere or made optically thin. 
\end{abstract}


\section{Introduction} \label{sec:intro}
Straddling the mass boundary between stars and planets lie brown dwarfs, substellar sources with masses ranging from $\sim13-75\,M_\mathrm{Jup}$ depending on their metallicity \citep{Saum96, Chab97} with temperatures ranging from $\sim250-3000$ K. This corresponds to spectral types of late-M, L, T, or Y \citep{Burg02a,Kirk05,Cush11} which are classified based on their optical or near-infrared (NIR) spectrum. Brown dwarfs continuously cool and contract throughout their lifetimes, evolving through the spectral sequence as they age due to their inability to sustain stable hydrogen burning.

Brown dwarfs can be split into roughly three main age groups: field dwarfs, low-gravity or young dwarfs, and old or subdwarfs. Field dwarfs are the base of the spectral classification scheme while the low-gravity dwarfs and subdwarfs expand the standard scheme. The low-metallicity brown dwarfs (subdwarfs) are distinguished from field sources by their unusually blue NIR $J-K$ colors \citep{Burg03c,Burg09a}, kinematics placing them in the galactic halo \citep{Dahn08, Burg08a, Cush09}, and spectral features of enhanced metal-hydride absorption bands (e.g. FeH), weak or absent metal oxides (TiO, CO, VO), and enhanced collisionally-induced H$\,_{2}$ absorption (\citealt{Burg03c} and references therein).

To explore the the information contained in the spectra of a source, we can use the data-driven method of atmospheric retrievals. Retrievals are a relatively new technique in brown dwarf science with only a small number of brown dwarfs having been retrieved to date \citep[e.g.][]{Line14b, Line15, Line17, Burn17,Burn21, Zale19, Kitz20, Gonz20}. The majority of these works have focused on examining cloud-free targets as they provide the simplest cases to interpret retrieval results. Additionally, brown dwarf retrievals have focused primarily on near-solar metallicity targets ([M/H]$=\pm0.5$) \citep[e.g.][]{Line17, Zale19, Kitz20, Gonz20, Burn21}. With subdwarfs being cloud-free due to the reduced condensates in their atmospheres, examining the impact low-metallicity has on retrieved parameters is a logical next step in building up complexity from the cloud-free solar metallicity sources. Thus, in an effort to explore the impact of metallicity on retrieved parameters and to characterize subdwarf atmospheric compositions, we examine the subdwarf SDSS J1256$-$0224.

In Section \ref{sec:literature} we present literature data on SDSS J1256$-$0224. We present updated semi-empirical fundamental parameters using distance-calibrated spectral energy distributions in Section \ref{FundParam1256}. Sections \ref{sec:RetrievalModel} and \ref{sec:modelselection} describes our retrieval framework and model selection. Results of our retrieval of SDSS J1256$-$0224 are discussed in Section \ref{sec:Retrieval_Rseults}.  Finally, Section~\ref{sec:Discussion} compares fundamental parameters from SED and retrieval methods to evolutionary models, how retrievals can help us to understand the origins of a source, how our results can inform low-metallicity models, and what can be learned from our rejected models.

\section{Literature Data on SDSS J1256$-$0224}\label{sec:literature}  
SDSS J125637.13$-$022452.4 (hereafter SDSS J1256$-$0224) was discovered by \cite{Siva09} as part of a search for L subdwarfs in the Sloan Digital Sky Survey data release 2 (SDSS DR2) spectral database. \cite{Burg09a} spectral typed it as a sdL3.5 and \cite{Zhang2017a} revised the spectral type to usdL3 following their scheme anchored to the work of \cite{Lepi07}. Equivalent widths of various spectral lines and spectral indices have been measured to aid in the classification of SDSS J1256$-$0224 in \cite{Scholz09}, \cite{Burg09a}, and \cite{Martin17}. There are currently five optical spectra available in the literature (SDSS: \citealt{Siva09}, VLT/FORS1: \citealt{Scholz09}, LDSS3: \citealt{Burg09a}, VLT/FORS2: \citealt{Lodi15}, and Palomar/DSpec: \citealt{Kirk16}) and 3 NIR spectra (SpeX prism: \citealt{Burg09a}, NIRSPEC: \citealt{Martin17}, FIRE echelle: \citealt{Gonz18}).

\textit{Gaia} eDR3 \citep{GaiaDR1,GaiaeDR3,Lind20} provides the most precise parallax and proper motions for SDSS J1256$-$0224 to date, with previous measurements by \cite{Siva09},\cite{Schi09}, and \textit{Gaia} DR2 \citep{GaiaDR1,GaiaDR2,Lind18}. \cite{Burg09a} measured a radial velocity and calculated $UVW$ velocities which placed SDSS J1256$-$0224 as an inner halo member. \cite{Gonz18} calculated updated $UVW$ velocities based on the \textit{Gaia} DR2 proper motions and parallax. Here we update the $UVW$ velocities and the tangential velocity using the \textit{Gaia} eDR3 parallax and proper motions listed in Table~\ref{tab:1256data}.

Fundamental parameters for SDSS J1256$-$0224 have been determined from fits to evolutionary model grids, fits to synthetic model atmosphere spectra and colors, and from distance calibrated spectral energy distributions (SEDs). In the literature the effective temperature of SDSS J1256$-$0224 has ranged from $T_\mathrm{eff}=1800-2600$~K based on the method used and data available at the time \citep{Siva09,Burg09a,Lodi15,Zhang2017a, Gonz18}, log\,$g$ from 5.0 to 5.5 \citep{Burg09a,Lodi15,Zhang2017a,Gonz18}, and [M/H] from $-2.0$ to $-1.0$ \citep{Schi09,Burg09a,Lodi15,Zhang2017a}. Values for SDSS J1256$-$0224 from the literature and those determined in this work are listed in Table~\ref{tab:1256data}.

\begin{deluxetable}{l c c}
\tabletypesize{\scriptsize}
\tablecaption{Properties of SDSS J1256$-$0224 \label{tab:1256data}}
\tablehead{\colhead{Property} & \colhead{Value} & \colhead{Reference}} 
  \startdata
  Spectral type & sdL3.5, usdL3 & 1, 2\\
  {[Fe/H]} & $-1.8 \pm 0.2$ & 2 \\\hline
  &\textbf{Astrometry}& \\ \hline
  R.A. & $12^h 56^m 37.16^s$ & 3 \\ 
  Decl. & $-02 ^\circ 24' 52''.2$ & 3 \\
  $\pi$ (mas) & $12.06 \pm 0.56$  & 4\\ 
  $\mu_\alpha$ (mas yr$^{-1}$) & $-510.85\pm0.73$ & 4\\          
  $\mu_\delta$ (mas yr$^{-1}$) & $-299.85\pm0.55$ & 4 \\
  $V_{r}$ (km s$^{-1}$) & $-126 \pm 10$ & 5 \\
  $V_\mathrm{tan}$ (km s$^{-1}$) & $232 \pm 10$ & 6 \\ 
  $U$ (km s$^{-1}$)\tablenotemark{a} & $-148\pm8.7$ & 6 \\            
  $V$ (km s$^{-1}$)\tablenotemark{a} & $-146.5\pm7.7$ & 6 \\ 
  $W$ (km s$^{-1}$)\tablenotemark{a} & $-163.7\pm9.1$& 6 \\ \hline
  &\textbf{Photometry} & \\ \hline
  PS $i$ (mag) & $19.47\pm0.04$ & 7 \\
  PS $z$ (mag) &$18.0\pm0.03$ & 7 \\
  PS $y$ (mag) &$17.54\pm0.01$& 7 \\
  SDSS $i$ (mag) & $19.41 \pm 0.02$ & 8\\
  SDSS $z$ (mag) & $17.71 \pm 0.02$ & 8\\
  2MASS $J$ (mag) & $16.10 \pm 0.11$ & 8\\
  2MASS $H$ (mag) & $15.79 \pm 0.15$ & 8\\
  2MASS $K_{s}$ (mag) & $<15.439$ & 8\\
  $H_\mathrm{MKO}$ (mag) & $16.078 \pm 0.016$ & 9\\
  $K_\mathrm{MKO}$ (mag) & $16.605 \pm 0.099$ & 10 \\
  WISE $W1$ (mag) & $15.214 \pm 0.038$ & 11\\
  WISE $W2$ (mag) & $15.106 \pm 0.098$ & 11\\
  WISE $W3$ (mag) & $<12.706$ & 11\\
  WISE $W4$ (mag) & $<8.863$ & 11\\\hline
  &\textbf{Parameters from SED}\tablenotemark{b} & \\ \hline 
  $L_\mathrm{bol}$ & $-3.59 \pm 0.04$ & 6 \\
  $T_\mathrm{eff}$ (K) & $2338\pm 56$ & 6\\
  Radius ($R_\mathrm{Jup}$) & $0.95\pm0.01$ & 6\\ 
  Mass ($M_\mathrm{Jup}$ )& $84\pm2$ & 6\\
  log\,$g$ (dex) & $5.37\pm0.01$ & 6, 10, 12\\                                                 
  Age (Gyr) & $5-10$ & 6, 10, 12\\
  Distance (pc) & $82.9\pm3.9$ & 6\\\hline
  &\textbf{Retrieved Parameters\tablenotemark{c}} & \\ \hline 
  log\,$g$ (dex) & $5.42\substack{+0.15 \\ -0.22}$ &  6\\ 
  \Lbol & $-3.59 \pm 0.01$ &  6\\ 
  \Teff (K) & $2558.37\substack{+168.43 \\ -137.97}$ &  6\\ 
  Radius ($R_\mathrm{Jup}$) & $0.79\substack{+0.11 \\ -0.10}$ &  6\\ 
  Mass ($M_\mathrm{Jup}$) & $65.14\substack{+21.41 \\ -20.61}$ &  6\\
  C/O & $\cdots$ & $\cdots$\\
  {[M/H]} & $-1.54\substack{+0.12 \\ -0.13}$ & 6\\
  \enddata
\tablenotetext{a}{UVW values are calculated in the local standard of rest frame in a left-handed coordinate system with U positive toward the Galactic center.}
\tablenotetext{b}{Using \cite{Saum08} low-metallicity (M/H$=-0.3$) evolutionary models, assuming an age of $0.5-10$~Gyrs.}
\tablenotetext{c}{\Lbol, \Teff, radius, mass, and {[M/H]} are not directly retrieved parameters, but are calculated using the retrieved $R^2/D^2$ and log\,$g$ values along with the predicted spectrum.}
\tablerefs{(1) \cite{Burg09a}, (2) \cite{Zhang2017a}, (3) \cite{Cutr03}, (4) \cite{GaiaDR1,GaiaeDR3,Lind20}, (5) \cite{Lodi15}, (6) This Paper, (7) \cite{Cham16}, (8) \cite{Adel08}, (9) \cite{Lawr12}, (10)\cite{Gonz18}, (11) \cite{Cutr12}, (12) \cite{Gonz19}.}
\tabletypesize{\small}
\end{deluxetable}

\section{Updated Fundamental Parameters of SDSS J1256$-$0224}\label{FundParam1256}

Fundamental parameters for SDSS J1256$-$0224 were determined using the technique of \cite{Fili15}, where we create a distance-calibrated SED using the spectra, photometry, and parallax.\footnote{SEDkit is available on GitHub at \url{https://github.com/hover2pi/SEDkit}. The Eileen branch was used for this work (also available at \url{https://github.com/ECGonzales/SEDkit/tree/master}).} The spectra (optical: LDSS3, NIR: FIRE echelle) and photometry used are the same as in \cite{Gonz19} with the \textit{Gaia} eDR3 parallax as the only update. The SED generation follows the exact procedure as done in \cite{Gonz18} using the same age and radius estimate. Our updated fundamental parameters are listed in Table~\ref{tab:1256data} and are consistent within 1$\sigma$ to the \cite{Gonz19} values. 

\section{The \textit{Brewster} Retrieval Framework \label{sec:RetrievalModel}} 
Our retrievals use the \textit{Brewster} framework \citep{Burn17} with a modified setup, similar to that of \cite{Gonz20}, in order to optimize for low metallicity atmospheres. A summary of the \textit{Brewster} framework with our modifications is discussed below. We differ from the \cite{Gonz20} setup by using an extended thermochemical grid that includes lower metallicities and higher temperatures. A more detailed description of \textit{Brewster} can be found in \cite{Burn17, Burn21}.

\subsection{The forward Model}
The \textit{Brewster} forward model uses the two-stream radiative transfer technique of \cite{Toon89}, including scattering, as first introduced by \cite{Mckay89} and subsequently used by e.g. \citet{Marl96, Saum08, Morl12}. We setup a 64 pressure layer (65 levels) atmosphere with geometric mean pressures from $\log P = -4$ to $2.3$ bars spaced in 0.1~dex intervals. To characterize the temperature in each layer we use the \cite{Madh09} parameterization which splits the atmosphere into three zones characterized by the following exponential functions:

\begin{equation}
\begin{aligned}
    P_0 < P < P_1&: P_0e^{\alpha_1(T-T_0)^{1/2}}\; (\mathrm{Zone\; 1}) \\
    P_1 < P < P_3&: P_2e^{\alpha_2(T-T_2)^{1/2}}\; (\mathrm{Zone\; 2}) \\
    P > P_3&: T = T_3\;\;\;\;\;\;\;\;\;\;\;\;\,  (\mathrm{Zone\; 3}). 
\end{aligned}
\label{eqn:madhu}
\end{equation}
$P_0$ and $T_0$ correspond to the pressure and temperature at the top of the atmosphere. At pressure $P_3$ with temperature $T_3$, the atmosphere becomes isothermal. In the general form, when $P_{2} > P_{1}$ a thermal inversion can occur, however, in this work we do not expect an inversion and therefore set $P_{2} = P_{1}$. $P_{0}$ is fixed in our model and continuity at the zonal boundaries is required, leaving us with five free parameters: $\alpha_{1}$, $\alpha_{2}$, $P_1$, $P_3$, and $T_3$.

\subsection{Gas Opacities} 
Optical depths due to absorbing gases for each layer are calculated using opacities sampled at a resolving power R~$=10000$ from the \cite{Free08,Free14} collection with the updated opacites from \cite{Burn17}. Line opacites in our pressure-temperature range are tabulated in 0.5~dex steps for pressure and for temperature in steps ranging from 20~K to 500~K as we move from 75~K to 4000~K, which is then linearly interpolated to our working pressure grid.

Due to the strong broadening of the D resonance doublets of \ion{Na}{1} ($\sim 0.59~\mu$m) and \ion{K}{1} ($\sim 0.77~\mu$m) in brown dwarf spectra, with line profiles detectable as far as $\sim$3000~cm$^{-1}$ from the line centre in T dwarfs \citep[e.g. ][]{Burr00,Lieb00,Marl02,King10}, a Lorentzian line profile is no longer an accurate representation for the line wings. Therefore for these two doublets we use line wing profiles based on the unified line shape theory \citep{nallard2007a, nallard2007b}. Tabulated line profiles (N. Allard, private communication) across temperatures of $500-3000$~K are calculated for the \ion{Na}{1} and \ion{K}{1} D1 and D2 lines broadened by collisions with H$_{2}$ and He for perturber (H$_{2}$ or He) densities up to $10^{20}$ cm$^{-3}$ with two collisional geometries considered for broadening by H$_{2}$. A Lorentzian profile with a width calculated from the same theory is used within 20~cm$^{-1}$ of the line center. Updated versions of these opacities exist \citep{nallard16,nallard19,Phil20}, however, we did not have access to them for this work. 

We include continuum opacities for H$_{2}$-H$_{2}$ and H$_{2}$-He collisionally induced absorption, using cross-sections from \citet{Rich12} and \citet{Saum12}, and include Rayleigh scattering due to H$_{2}$, He and CH$_{4}$ but we neglect the remaining gases. Free-free continuum opacities are included for H$^{-}$ and H$_2^{-}$ as well as bound-free continuum opacity for H$^{-}$ \citep{Bell+1987,bell1980,john1988}, which are influenced by the H$^-$ metallicity and determined from the thermochemical equilibrium grid (see Section~\ref{sec:gas_abundances}). We assume the atmosphere is dominated by H$_2$ and He, with proportions of (0.84H$_2+$ 0.16He) based on Solar abundances. H$_2$ and He are assumed to make up the remainder of the gas in a layer after including the retrieved gases, neutral H, H$^{-}$, and electrons. Neutral H, H$^{-}$, and electrons are drawn from the thermochemical equilibrium grids discussed later in this section.

\subsection{Determining Gas Abundances \label{sec:gas_abundances}} 
As done in \cite{Gonz20}, we use two methods to determine gas abundances-- (1) uniform-with-altitude mixing ratios and (2) thermochemical equilibrium. The uniform-with-altitude method, also known as ``free'' retrievals, enables us to directly retrieve individual gas abundances. However, the simplicity of this method can not capture important variations with altitude that are expected for some species which can vary by several orders of magnitude in the photosphere. These variations are expected to have a large contribution to the flux we observe e.g. metal-oxides and metal-hydrides of SDSS J14162408$+$1348263A (hereafter SDSS J1416$+$1348A) and the alkalies for SDSS J14162408$+$1348263B (hereafter SDSS J1416$+$1348B) in \cite{Gonz20} and the metal-hydrides of SDSS J1256$-$0224. We would prefer to freely retrieve abundances that vary with altitude; however, the resultant large number of parameters to solve for in this approach is computationally prohibitive, also ill posed. 

The thermochemical equilibrium method aims to address this issue by instead retrieving [Fe/H] and the carbon-to-oxygen (C/O) ratio. Gas fractions in each layer are then pulled from thermochemical equilibrium abundance tables as a function T, P, [Fe/H], and the C/O ratio along with the thermal profile of a given state-vector. We use thermochemical equilibrium grids calculated using the NASA Gibbs minimization CEA code \citep{McBr94}, based on previous thermochemical models \citep{Fegl94,Fegl96,Lodd99,Lodd02,Lodd02b,Lodd10,Lodd06,Viss06,Viss10a,Viss12,Moses12,Moses13} and recently utilized to explore gas and condensate chemistry over a range of conditions in substellar atmospheres \citep{Morl12,Morl13,Skem16,Kata16,Wake17,Gonz20,Burn21,Ghar21}. The chemical grids in this work determine equilibrium abundances of atmospheric species over pressures ranging from 1 microbar to 300 bar, temperatures between $500-6000$~K, metallicities ranging from $-2.5 < [\mathrm{Fe/H}] < +3.0$, and C/O abundance ratios of $0.25$ to $2.5$x the solar abundance (\textit{bulk} C/O$=0.46$ based on \citealt{Lodd10} abundances) and are extensions of those in \cite{Sonora}.

\subsection{Cloud Model \label{sec:cloudmodel}}
We use the same simple cloud model as done in \cite{Burn17} and \cite{Gonz20}, where the cloud is parameterized as a ``deck'' or ``slab''. Both cloud types are defined where opacity due to the cloud is distributed among layers in pressure space and the optical depth defined as either grey or as a power-law ($\tau = \tau_0\lambda^\alpha$, where $\tau_0$ is the optical depth at 1~$\mu$m). As done in \cite{Gonz20}, the single scattering albedo is set to zero thereby assuming an absorbing cloud.

The deck cloud is defined to always become optically thick at some pressure $P_{top}$, such that we only see the cloud top and the vertical extent of the cloud at pressures lower than $P_{top}$. There are three parameters for the deck cloud: (1) a cloud top pressure $P_{top}$, the point at which the cloud passes $\tau=1$ (looking down), (2) the decay height $\Delta \log P$, over which the optical depth falls to lower pressures as $d\tau/dP \propto \exp((P-P_{deck}) / \Phi)$ where $\Phi = (P_{top}(10^{\Delta \log P} - 1))/(10^{\Delta \log P})$, and (3) the cloud particle single-scattering albedo. At pressures $P >P_{top}$, the optical depth increases following the decay function until $\Delta \tau_{layer} = 100$. Deep below the cloud top there is essentially no atmospheric information as the deck cloud can rapidly become opaque with increasing pressure. Therefore, we stress that the pressure-temperature (P-T) profile below the deck cloud top extends the gradient at the cloud top pressure. 

The slab cloud differs from the deck as it is possible to see the bottom of the slab. The slab parameters include (1) and (3) as done for deck. Because it is possible to see to the bottom of the slab, parameter (2) is now a physical extent in log-pressure ($\Delta \log P$). We also include an additional parameter for determining the total optical depth of the slab at 1$\mu$m ($\tau_{cloud}$), bringing the total number of parameters to 4. The optical depth is distributed through the slab's extent as $d\tau / dP \propto P$ (looking down), reaching its total value at the highest pressure (bottom) of the slab. The slab can have any optical depth in principle, however, we restrict our prior as $0.0 \leq \tau_{cloud} \leq 100.0$.

If the deck or slab cloud is non-grey, an additional parameter for the power ($\alpha$) in the optical depth is included. 

\subsection{Retrieval Model}\label{sec:Retmodelsetup} 
We use the EMCEE package \citep{emcee} to sample posterior probabilities as done in \cite{Gonz20} with the priors we used shown in Table \ref{tab:Priors}. For the SDSS J1256$-$0224 retrievals we extend the mass and temperature priors to allow for surface gravities that encompass the possible ranges for subdwarfs. We use the distance-calibrated SpeX prism spectra (output from generating our SED using the \textit{Gaia} DR2 parallax as the eDR3 parallax was not available at the start of this work) trimmed to the $1.0-2.5\,\mu$m region and set the distance to 10~pc with the correspondingly scaled uncertainty for our retrieval as done in \cite{Gonz20}.

\begin{deluxetable*}{l c c c c}
\tablecaption{Data used for SDSS J1256$-$0224 retrieval models\label{tab:retdata}} 
\tablehead{\colhead{Spectrum} &\colhead{Obs. date} &\colhead{Reference} & \colhead{Distance (pc)} & \colhead{Reference}}
  \startdata
  Spex prism\tablenotemark{a} & 2005-03-23 & \cite{Burg09a} & $79.7\pm4.6$\tablenotemark{b}& \textit{Gaia} DR2
  \enddata
  \tablenotetext{a}{Distance-calibrated output from SEDkit.}
  \tablenotetext{b}{Because we used the 10 pc distance-calibrated spectrum, we set the distance in our retrievals to 10 pc instead of the true distance and scale the uncertainty accordingly. Therefore we set the distance to $10\pm0.58$ pc.}
  \tablecomments{We use the \textit{Gaia} DR2 parallax as the eDR3 value was not available at the start of this work.}
\end{deluxetable*}

\begin{deluxetable*}{l c c}
\tablecaption{Priors for SDSS J1256$-$0224 retrieval models\label{tab:Priors}} 
\tablehead{\colhead{Parameter} &\phm{stringzzzzzzzz} &\colhead{Prior}}
  \startdata
  gas volume mixing ratio && uniform, log $f_{gas} \geq -12.0$, $\sum_{gas} f_{gas} \leq 1.0$ \\
  thermal profile ($\alpha_{1}, \alpha{2}, P1, P3, T3$) && uniform, $0.0\, \mathrm{K} < T < 6000.0\, \mathrm{K}$\\ 
  scale factor ($R^2/D^2$) && uniform, $0.5\,R_\mathrm{Jup} \leq\, R\, \leq 2.0\,R_\mathrm{Jup}$ \\
  gravity (log\,$g$)&& uniform, $1\,M_\mathrm{Jup} \leq\; gR^2/G\; \leq 100\,M_\mathrm{Jup}$ \\
  cloud top\tablenotemark{a} && uniform, $-4 \leq \mathrm{log}\, P_{CT} \leq 2.3$\\ 
  cloud decay scale\tablenotemark{b} && uniform,$0< \mathrm{log}\,\Delta\, P_{decay}<7$\\
  cloud thickness\tablenotemark{c} && uniform, log\,$P_{CT} \leq\,$log $(P_{CT}+\Delta P)\, \leq2.3$\\
  cloud total optical depth at $1\mu$m && uniform,  $0.0 \geq \tau_{cloud} \geq 100.0$ \\
  single scattering albedo ($\omega_0$) && uniform, $0.0 \leq \omega_0 \leq 1.0$\\
  wavelength shift && uniform, $-0.01 < \Delta \lambda <0.01 \mu$m \\
  tolerance factor && uniform, log($0.01 \times min(\sigma_{i}^2)) \leq b \leq$ log$(100 \times max(\sigma_{i}^2)) $\\
  \enddata
  \tablenotetext{a}{For the deck cloud this is the pressure where $\tau_{cloud} = 1$, for a slab cloud this is the top of the slab.}
  \tablenotetext{b}{Decay height above the $\tau_{cloud} = 1.0$ level only for deck cloud.}
  \tablenotetext{c}{Thickness and $\tau_{cloud}$ only retrieved for slab cloud.}
\end{deluxetable*}

For each model for SDSS J1256$-$0224, we initialize 16 walkers per parameter in a tight gaussian for the gases, surface gravity, $\Delta\lambda$ (the wavelength shift between the model and data), and the scale factor ($R^2/D^2$). We center our gases around the approximate solar composition equilibrium chemistry values for gas volume mixing ratios and the surface gravity is initialized centered around the evolutionary model value from the SED analysis. For our tolerance parameter, we use a flat distribution across the entire prior range. Lastly, the cloud top pressure and power-law parameters are initialized as tight Gaussians, while we use a flat prior across the entire range for optical depth, albedo, and cloud thickness. For our thermal profile we use the \cite{Saum08} \Teff$=1700$~K log\,$g=5.0$ model to initialize $\alpha_{1}$, $\alpha_{2}$, $P_1$, $P_3$, and $T_3$. To be sure of convergence, each model is run for at least 50 times the autocorrelation length with the EMCEE chain having between 90,000-120,000 iterations.

In this work we retrieved abundances for the following gases: H$_2$O, CO, CO$_2$, CH$_4$, TiO, VO, CrH, FeH, K, and Na, which we will refer to as the ``full gas set''. As done in prior works \citep{Line15, Burn17, Gonz20} we tie K and Na together as a single element in the state-vector assuming a Solar ratio taken from \cite{Aspl09}. The H$^-$ bound-free and free-free continuum opacities are included to account for the possibility of the profile going above 3000K in the photosphere. These are set based on temperature and pulled from the chemical grid. Various cloud parameterizations are tested building up from cloudless to the 4 parameter power-law slab cloud model. We have the capability of testing more complex cloud models (e.g. specific species, multi-clouds, etc. see \citealt{Burn21}), however we do not test these as they are not warranted based on our simple clouds. Prior to moving from cloud-free to cloudy models, we tested the impact of metallicity for determining neutral H, H$^{-}$, and electron abundances used for the continuum opacity calculations. For completeness, we also tested the uniform-with-altitude and thermochemical equilibrium gas abundance methods on the winning model. As discussed in Section~\ref{sec:Retrieval_Rseults}, we were unable to strongly constrain CO, CO$_2$, CH$_4$, TiO, VO, K, and Na and therefore additionally tested models only including H$_2$O, CrH, and FeH on our top 4 models. We will refer to these as the ``reduced gas set models''.

\section{Model Selection}\label{sec:modelselection}
The four key aspects tested with our selection of retrieval models were (1) impact of ion metallicity, (2) cloud parameterization, (3) gas abundance method, and (4) gas species included. To rank our results, we used the Bayesian Information Criterion (BIC), where the lowest BIC is preferred. In Table~\ref{tab:1256RetrievalModels} we rank the models with increasing $\Delta$BIC from the ``winning'' model. We use the following intervals from \cite{Kass95} for selecting between two models, with evidence against the higher BIC as:

\begin{itemize}
\setlength\itemsep{-0.5em}
  \item[] $0  < \Delta$BIC $< 2$: no preference worth mentioning;
  \item[] $2  < \Delta$BIC $< 6$: positive;
  \item[] $6  < \Delta$BIC $< 10$: strong;
  \item[] $10 < \Delta$BIC: very strong.
\end{itemize}

Cloud parameterization is explored in our retrievals starting with the cloud-free model and building up in complexity to the most complex slab cloud model. Before starting cloudy retrievals because we expect a subsolar metallicity for SDSS J1256$-$0224, we tested the impact of choosing metallicities ranging from $-2.5\leq\mathrm{[Fe/H]}_{ion}\leq0.0$ when determining the neutral H, H$^{-}$, and electron (ion) abundances, which are used for the continuum opacity calculations. These ion metallicities were chosen from a chemical grid assuming a Solar C/O ratio.  When using the full gas set, for metallicities below [Fe/H]$_{ion}=-1.5$ the cloud-free models were indistinguishable from one another. We therefore ran models using [Fe/H]$_{ion}=-1.5,-2,$ and $-2.5$ for each of our cloudy models. 

As SDSS J1256$-$0224 could have a subsolar C/O ratio, we also ran a cloud-free, [Fe/H]$_{ion}=-1.5$ ion metallicity model where the ion abundances were pulled from a chemical grid with a C/O ratio of $0.25x$ Solar. We found that this model was indistinguishable from the cloud-free, [Fe/H]$_{ion}=-1.5$, Solar C/O model and therefore used ion abundances from the Solar C/O ratio chemical grid for the cloudy models. 

Additionally, before moving to the cloudy models we tested the alternate gas abundance method (thermochemical equilibrium) on the ``winning'' uniform-with-altitude model. In this case since the [Fe/H]$_{ion}=-1.5,-2,$ and $-2.5$ full gas set models were indistinguishable, we only tested thermochemical equilibrium on the [Fe/H]$_{ion}=-2$ model. 

Lastly, we were unable to strongly constrain some gases included in our models. To test the impact of these gases we ran [Fe/H]$_{ion}=-1.5,-2,$ and $-2.5$ cloud-free models including only the constrained gases (H$_2$O, CrH, and FeH). As these reduced gas set models produced significantly better $\Delta$BIC values and model spectra similar to full gas set models, we conclude that the additional gases initially included are not constrained.

\section{Retrieval Model of SDSS J1256$-$0224}\label{sec:Retrieval_Rseults}

Table \ref{tab:1256RetrievalModels} lists all models tested for SDSS J1256$-$0224 with their corresponding $\Delta$BIC values. The cloud-free uniform-with-altitude ion metallicity [Fe/H]$_{ion}=-1.5$ reduced gas set model best fits SDSS J1256$-$0224. When comparing the BIC to the full gas set version of this model, the reduced gas set model is greatly preferred. The improved BIC for the reduced gas set model is likely due to the fewer parameters, as the spectral fits are nearly identical to those in the full gas set model.

We, however find that the cloud-free uniform-with-altitude [Fe/H]$_{ion}=-2$ reduced gas set model and the cloud-free uniform-with-altitude [Fe/H]$_{ion}=-1.5$ drawn from the C/O$=0.25x$ Solar chemical grid reduced gas set model (with a $\Delta$BIC$=1.6$ and $\Delta$BIC$=0.4$ respectively) are indistinguishable from the ``winning'' model and therefore provides similar quality fits to SDSS J1256$-$0224's spectroscopic features. As the P-T profiles and spectra of these indistinguishable models are very similar to the winning model, we will only discuss the winning model here. Discussion of the indistinguishable models in comparison to the ``winning'' model can be found in Section \ref{sec:indistinguishable_model_comp}.

\begin{deluxetable*}{l c c c c}
\tablecaption{SDSS J1256$-$0224 Retrieval Models with $\Delta$BIC \label{tab:1256RetrievalModels}} 
\tablehead{\colhead{Cloud Model} & Gas Method\tablenotemark{a} &\colhead{Ion Metallicity ([Fe/H])} & \colhead{Number of Parameters} & \colhead{$\Delta$BIC}}
  \startdata
  \multicolumn{5}{c}{Reduced Gas Set Models\tablenotemark{b}} \\ \hline
  Cloud Free           & uniform & -1.5  & 12 & 0 \\ 
  Cloud Free           & uniform & -1.5\tablenotemark{c} & 12 & 0.4 \\ 
  Cloud Free           & uniform & -2.0  & 12 & 1.6 \\ 
  Cloud Free           & uniform & -2.5  & 12 & 4.9 \\ \hline 
  \multicolumn{5}{c}{Full Gas Set Models\tablenotemark{d}} \\ \hline
  Cloud Free           & uniform & -1.5  & 18 & 36.4 \\
  Cloud Free           & uniform & -1.5\tablenotemark{c} & 18 & 36.7 \\ 
  Cloud Free           & uniform & -2.5  & 18 & 37.0 \\
  Cloud Free           & uniform & -2.0  & 18 & 37.0 \\ \hline
  Cloud Free           & uniform & -1.0  & 18 & 45.6 \\
  Grey Deck cloud      & uniform & -1.5 & 21 & 56.1 \\
  Grey Deck cloud      & uniform & -2.0 & 21 & 56.5 \\
  Grey Deck cloud      & uniform & -2.5 & 22 & 59.4 \\
  Grey Slab cloud      & uniform & -2.0 & 22 & 59.9 \\  
  Power-law Deck cloud & uniform & -1.5 & 22 & 62.1 \\
  Grey Slab cloud      & uniform & -1.5 & 22 & 62.2 \\
  Grey Slab cloud      & uniform & -2.5 & 22 & 62.5 \\
  Power-law Deck cloud & uniform & -2.0 & 22 & 62.8\\
  Power-law Deck cloud & uniform & -2.5 & 22 & 65.9 \\
  Power-law Slab cloud & uniform & -2.0 & 23 & 67.4 \\
  Power-law Slab cloud & uniform & -2.5 & 23 & 68.3 \\  
  Power-law Slab cloud & uniform & -1.5 & 23 & 68.6 \\
  Cloud Free           & CE  & -2.0  & 11 & 102.5 \\
  Cloud Free           & uniform & Solar & 18 & 129.2 \\
  \enddata
  \tablenotetext{a}{Method used to determine gas abundances. uniform = uniform-with-altitude mixing, CE = Chemical Equilibrium.}
  \tablenotetext{b}{H$_2$O, CrH, and FeH only.}
  \tablenotetext{c}{Ion abundances drawn from chemical grid with C/O=$0.25x$ Solar, where \textit{bulk} Solar C/O=0.46 based on \cite{Lodd10} abundances.}
  \tablenotetext{d}{H$_2$O, CO, CO$_2$, CH$_4$, TiO, VO, CrH, FeH, and Na+K.}
\end{deluxetable*}

\subsection{Pressure-Temperature Profile and Contribution Function\label{sec:PT_cont_deck}}
\begin{figure*}
\gridline{\fig{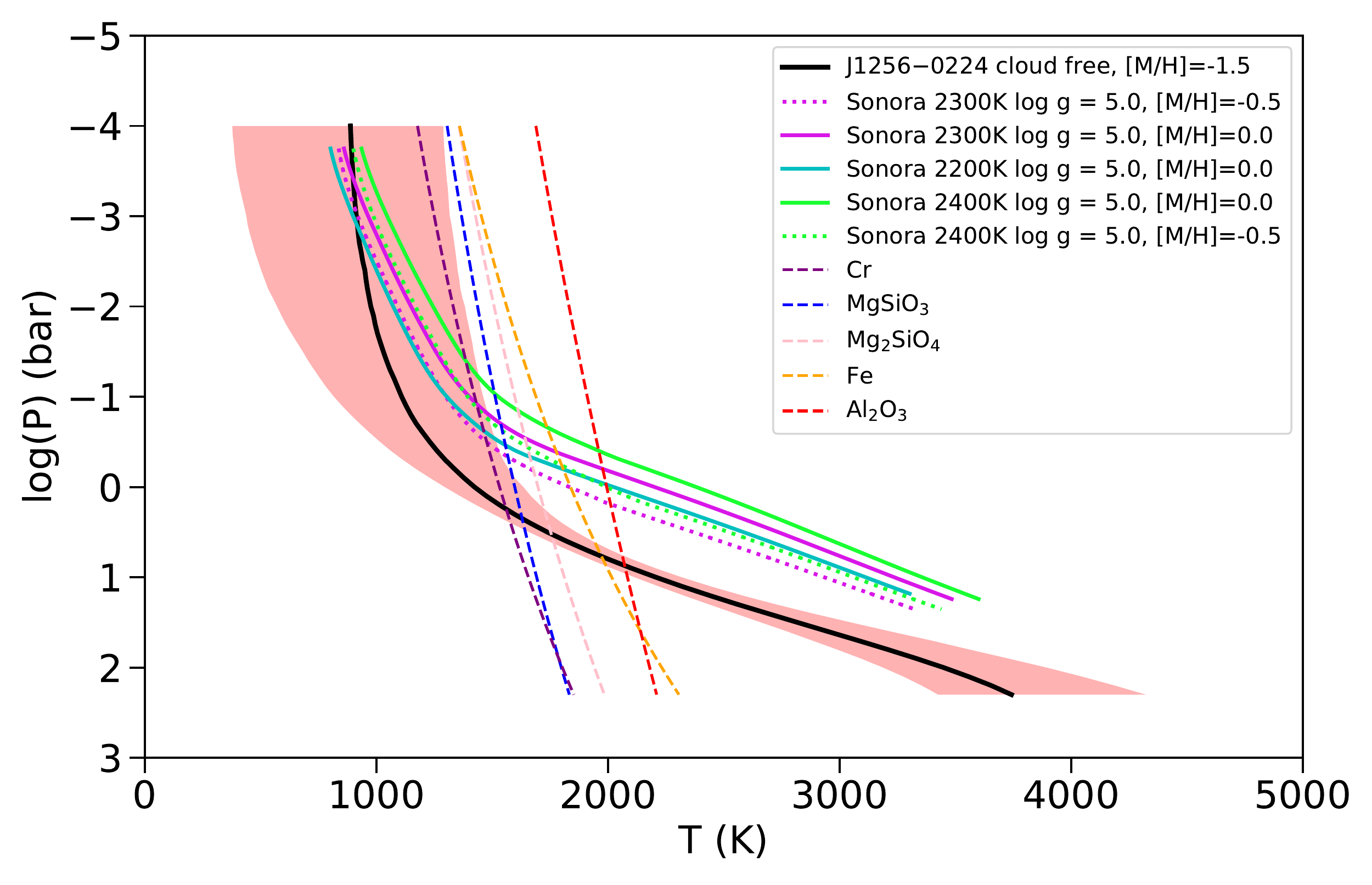}{0.5\textwidth}{\large(a)} 
          \fig{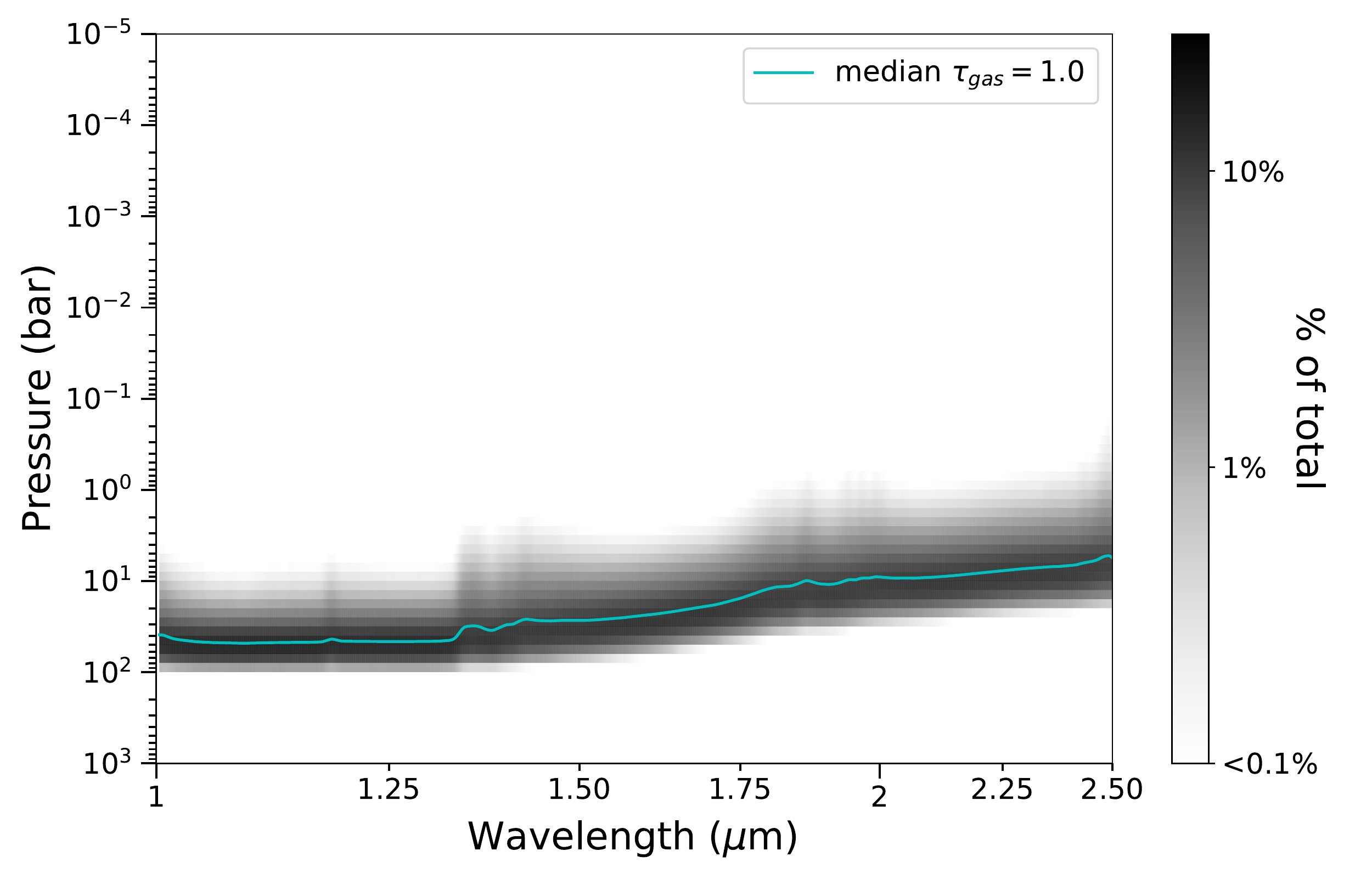}{0.5\textwidth}{\large(b)}} 
\caption{(a) Retrieved Pressure-Temperature Profile of SDSS J1256$-$0224 (median in black, 1 $\sigma$ pink shading) compared to Sonora cloudless solar and low-metallicity model profiles (neon green, teal, and purple). Colored dashed lines are condensation curves for the listed species. (b) The associated contribution function for this model, with median gas $\tau=1$ contribution over plotted in aqua.}
\label{fig:1256_profile_contribution}
\vspace{0.5cm} 
\end{figure*}

Figure~\ref{fig:1256_profile_contribution}(a) shows the retrieved P-T profile for the ``winning'' model for SDSS J1256$-$0224. Overplotted are the Sonora \citep{Sonora} profiles for solar-metallicity (solid lines) and [M/H]$=-0.5$ (dotted lines) at a log$\,g=5.0$ with temperatures from $2200-2400$~K which bracket the semi-empirical and retrieved $T_\mathrm{eff}$. P-T profiles from the Sonora grid currently only are provided for metallicities as low as [M/H]$=-0.5$. In the photosphere ($\sim5-75$~bars, see panel b) and deeper, the slope of our P-T profile matches that of the Sonora grid model profiles. In the upper atmosphere, at pressures lower than 0.1 bar, our retrieved profile becomes nearly isothermal unlike the Sonora profiles.

The contribution function for this model is shown in Figure \ref{fig:1256_profile_contribution}(b) with the $\tau=1$ gas contribution in aqua. The contribution function in a layer is defined as

\begin{equation}
\begin{aligned}
    C(\lambda, P) =\frac{B(\lambda,T(P))\int_{P_1}^{P_2}d\tau}{\exp{\int_0^{P_2}d\tau}}
\end{aligned}
\label{eqn:madhu}
\end{equation}

where $B(\lambda,T(P))$ is the Planck function, the pressure at the top of the atmosphere is zero, $P_1$ is the pressure at the top of the layer, and $P_2$ is the pressure at the bottom of the layer. The bulk of the flux contributing to the observed spectrum, corresponding to the photosphere, comes from roughly 5 to 75 bars, reaching to deeper layers than the previously retrieved L dwarfs in \cite{Burn17} ($1-10$ bars) and \cite{Gonz20} ($1-18$ bars).

\subsection{Retrieved gas abundances and derived properties}
\begin{figure*}
  \centering
   \includegraphics[scale=.4]{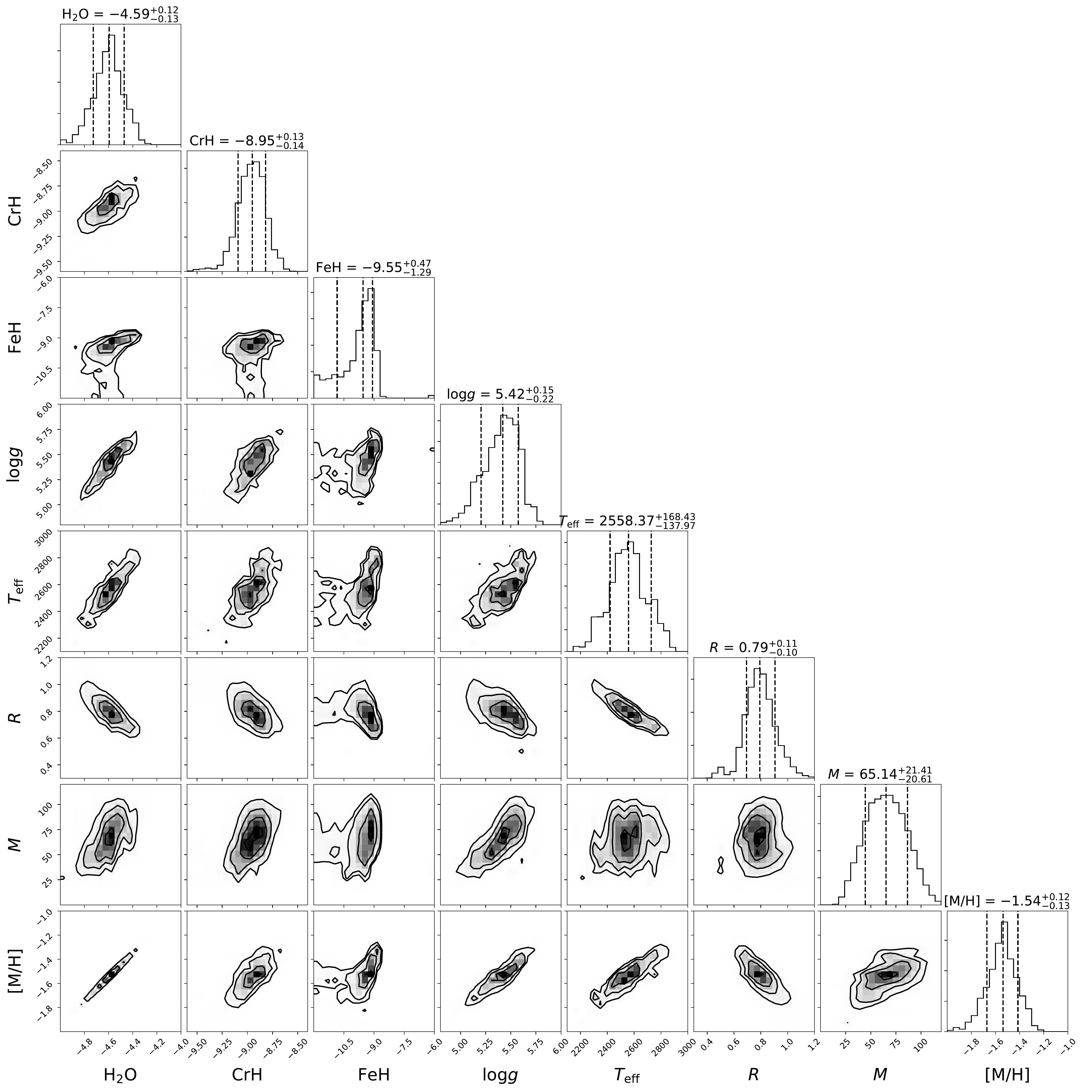} 
\caption{SDSS J1256$-$0224 ``winning'' model posterior probability distributions for the retrieved parameters and extrapolated parameters. 1D histograms of the marginalized posteriors are shown along the diagonals with 2D histograms showing the correlations between the parameters. The dashed lines in the 1D histograms represent the 16\textsuperscript{th}, 50\textsuperscript{th}, and 84\textsuperscript{th} percentiles, with the 68\% confidence interval as the width between the 16\textsuperscript{th} and 84\textsuperscript{th} percentiles. Parameter values listed above are shown as the median~$\pm1\sigma$. Gas abundances are displayed as log$_{10}$(X) values, where X is the gas. \Teff, radius, mass, and {[M/H]} are not directly retrieved parameters, but are calculated using the retrieved $R^2/D^2$ and log\,$g$ values along with the predicted spectrum. Our [M/H] is relative to Solar.}
\label{fig:1256_gascorner}
\vspace{0.5cm} 
\end{figure*}

\begin{deluxetable}{l c}
\tablecaption{Retrieved Gas Abundances and Derived Properties for SDSS J1256$-$0224 \label{tab:1256_Corner_values}} 
\tablehead{\colhead{Parameter}\phm{stringssssssssssssssssss} & \colhead{Value}}
  \startdata
  \multicolumn{2}{c}{Retrieved} \\\hline
  H$_2$O & $-4.59\substack{+0.12 \\ -0.13}$\\ 
  CO & <$-5.63$\\ 
  CO$_2$ & <$-4.79$\\
  CH$_4$ & <$-6.14$\\
  TiO & <$-9.75$\\ 
  VO & <$-9.92$\\ 
  CrH & $-8.95\substack{+0.13 \\ -0.14}$\\
  FeH & $-9.55\substack{+0.47 \\ -1.29}$\\
  Na+K & <$-8.64$\\ 
  log\,$g$ (dex) & \phm{+}$5.42\substack{+0.15 \\ -0.22}$ \\ \hline
  \multicolumn{2}{c}{Derived}  \\ \hline
  \Lbol & $-3.59 \pm 0.01$ \\ 
  \Teff (K) & $2558.37\substack{+168.43 \\ -137.97}$ \\ 
  Radius ($R_\mathrm{Jup}$) & \phm{+}$0.79\substack{+0.11 \\ -0.10}$ \\ 
  Mass ($M_\mathrm{Jup}$) & \phm{+}$65.14\substack{+21.41 \\ -20.61}$ \\
  {C/O} & $\cdots$ \\ 
  {[M/H]}\tablenotemark{a} & $-1.54\substack{+0.12 \\ -0.13}$ \\
  \enddata
  \tablenotetext{a}{Atmospheric value. Derived from only H$_2$O, CrH, and FeH abundances.}
  \tablecomments{Molecular abundances are fractions listed as log values. Upper limits for gases not included in this model come from the cloud-free, [M/H]$=-1.5$ full gas set model. For unconstrained gases, 1$\sigma$ confidence is used to determine upper limits.} 
\end{deluxetable}

Figure~\ref{fig:1256_gascorner} shows the posterior probability distributions for the retrieved gas abundances and surface gravity, the derived quantities for radius, mass, and atmospheric metallicity, and the extrapolated \Teff. Table~\ref{tab:1256_Corner_values} provides the values from the corner plot for ease of reading and to present upper limits for unconstrained parameters. 

Our derived radius is determined by using the retrieved scale factor and the parallax measurement. The mass is then calculated using the derived radius and the retrieved log\,$g$. To get the extrapolated \Teff we use the radius and integrate the flux from the resultant forward model spectrum across $0.7-20$~$\mu$m. The atmospheric metallicity is derived using the following equations:
\begin{equation}
    f_{H_2}=0.84(1-f_{gases})
\end{equation}
\begin{equation}
    N_{H}=2f_{H_2}N_{tot}
\end{equation}
\begin{equation}
    N_{element}=\sum_{molecules}  n_{atom}f_{molecule}N_{tot}
\end{equation}
\begin{equation}
    N_{M}=\sum_{elements} \frac{N_{element}}{N_H}
\end{equation}
where $f_{H_2}$ is the H$_2$ fraction, $f_{gases}$ is the total gas fraction containing all constrained gases, $N_{H}$ is the number of neutral hydrogen atoms, $N_{tot}$ is the total number of gas molecules, $N_{element}$ is the number of atoms for the element of interest, and $n_{atom}$ is the number of atoms of that element in a molecule (e.g. 2 for oxygen in CO$_2$). Thus the final value of [M/H] is
\begin{equation}
    [M/H]=log\frac{N_M}{N_{Solar}}
\end{equation}
where $N_{Solar}$ is calculated as the sum of the solar abundances (from \citealt{Aspl09}) relative to H. We do not account for gases that are invisible (e.g. Nitrogen) in our atmospheric metallicity calculation.

From our initial full gas set model, we were only able to constrain abundances for H$_2$O, CrH, and FeH. One likely explanation is the low metallicity of SDSS J1256$-$0224. As subdwarfs have weak or absent metal oxides (TiO, VO, CO) along with enhanced collisionally-induced $H_2$ absorption, which can aid in the reduction of CO in the $K$ band (\citealt{Burg09a} and references therein), these species may not be in the atmosphere of SDSS J1256$-$0224. Another explanation could be that the signal-to-noise (SNR) is not high enough in particular regions of the spectrum for some gas species. This could be a possible cause for the unconstrained CO abundance in particular, where \cite{Gonz18} saw weak CO absorption in the FIRE echelle in the $K$ band spectrum of SDSS J1256$-$0224 that was undetectable in the SpeX prism spectrum. Both cases will be discussed in Section~\ref{sec:Discussion}. We derive a metallicity of [M/H]$=-1.54\substack{+0.12 \\-0.13}$ and find that it agrees with the \cite{Zhang2017a} value. As we do not constrain any carbon-bearing species, we cannot determine a C/O ratio for SDSS J1256$-$0224. As a subdwarf, we do not expect to detect a large fraction of carbon in the atmosphere due to the low amount of carbon available in the natal gas at the time of formation. With these species excluded from our metallicity calculations, our metallicity is likely overestimated (lower than the true value). A comparison of our constrained parameters to those expected from chemical grids are discussed in Section~\ref{sec:spectrum_and_comp}.

\subsection{Retrieved Spectrum and Composition\label{sec:spectrum_and_comp}}
\begin{figure*}
\centering
 \gridline{\fig{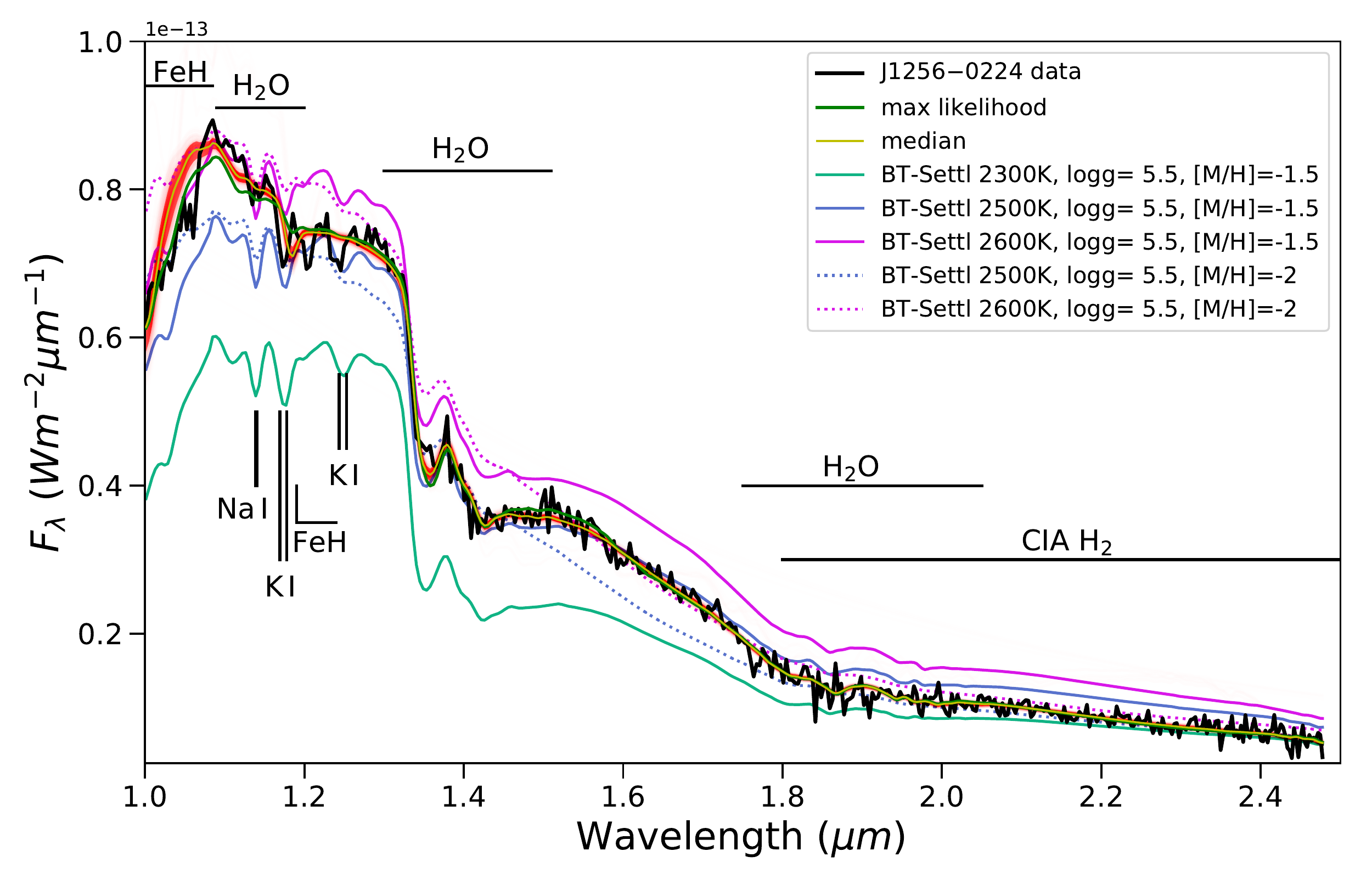}{0.5\textwidth}{\large(a)}
           \fig{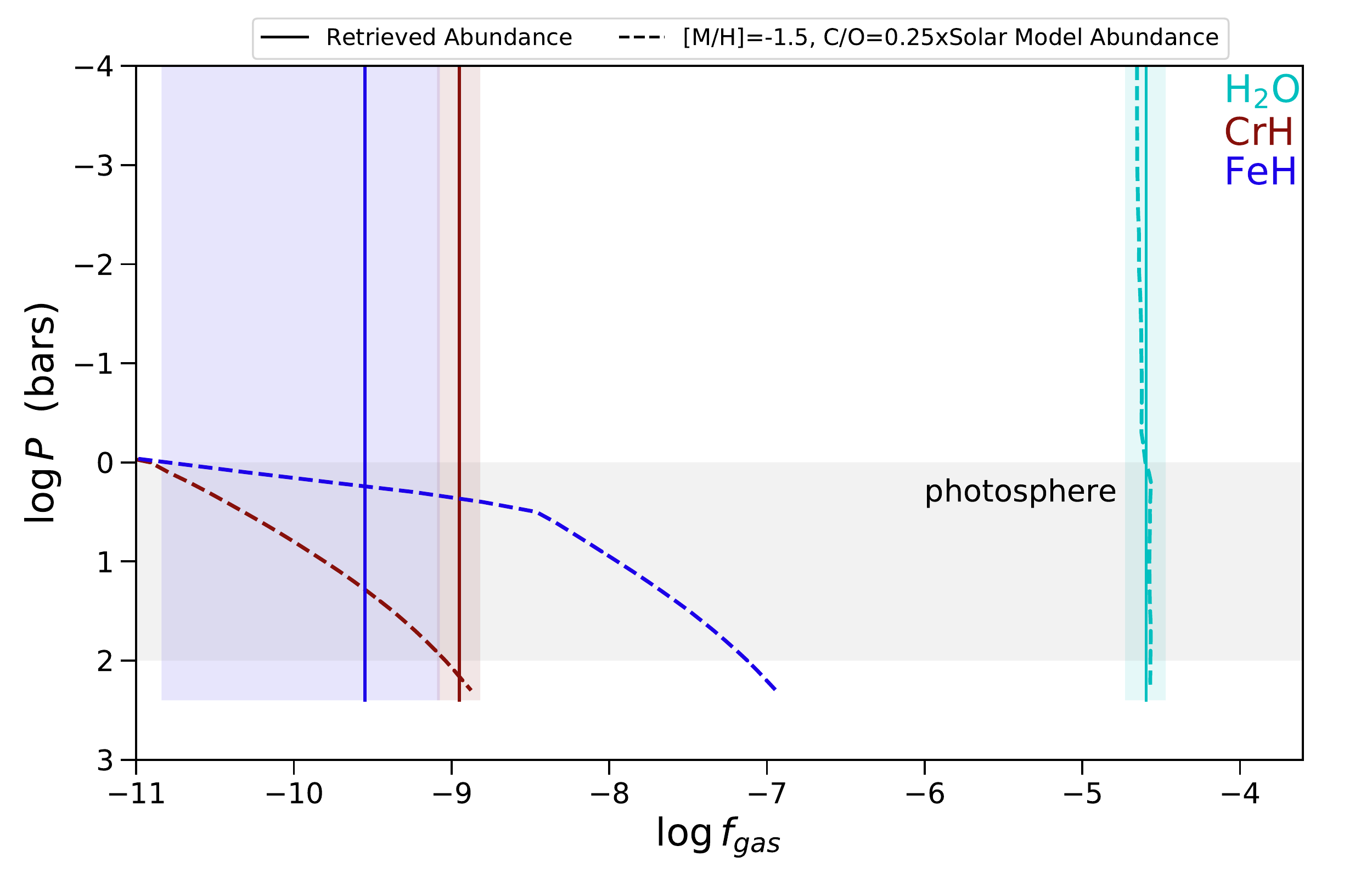}{0.5\textwidth}{\large(b)}}
\caption{(a) Retrieved forward model spectra of SDSS J1256$-$0224 for the ``winning'' cloud-free, ion metallicity [M/H]=-1.5 reduced gas set model. The maximum likelihood spectrum is shown in dark green, the median spectrum in yellow, and 500 random draws from the final 2000 samples of the EMCEE chain in red. The SpeX prism data is shown in black. For comparison the BT-Settl low metallicity grid models bracketing SED-derived and retrieval-derived \Teff range are over-plotted (log$\,g=5.5$, [M/H]$=-1.5,-2$ (solid/dotted lines), \Teff$=2300-2600$K (teal, blue, and purple)). (b) Retrieved uniform-with-altitude mixing abundances for constrained gases compared to abundances expected for a [M/H]$=-1.5$ and C/O$=0.25x$ Solar model. Bulk Solar C/O$=0.46$. The approximate location of the photosphere is shown in gray. \textbf{Gas abundances are shown as log\;$f_{gas}$ and are unitless.}}
\label{fig:1256_spectrum_vmr}
\vspace{0.5cm} 
\end{figure*}

Figure \ref{fig:1256_spectrum_vmr}(a) compares the fit of our retrieved spectrum and the BT-Settl\footnote{\url{https://phoenix.ens-lyon.fr/Grids/BT-Settl/AGSS2009/}.\\ Retrieved from \url{http://svo2.cab.inta-csic.es/theory/newov2/index.php?models=bt-settl-agss}} grid models (\citealt{Alla12}, scaled to our median retrieved scale factor) to the observed SpeX prism data. The BT-Settl 2500~K, log$\,g=5.5$, [Fe/H]$-1.5$ model provides the overall best fit to the observed data, however it has difficulties fitting the $Y$ and $J$ bands. Our retrieved spectrum does a better job than the BT-Settl models at fitting the spectrum overall, however we also struggle at fitting the features in the $Y$ and $J$ bands. The largest issue in these bands is the inability of our model to simultaneously fit the slope from $\sim1-1.1\,\mu$m, which is highly influenced by the pressure broadened wings of the $0.77\mu$m \ion{K}{1} doublet, and the narrow \ion{Na}{1} and \ion{K}{1} doublets. We see the fit to the 1.25 \ion{K}{1} in the $J$ band appears to be shifted to the right of the feature, due to it being blended with a CrH feature nearby. Thus we caution the reader in trusting the retrieved CrH abundance due to this blend. These issues are likely due to how pressure broadening is treated in the opacity models for these lines. Additionally, we are unable to fit the FeH feature in the $J$-band. This was seen in the retrieval of SDSS J1416$+$1348A \citep{Gonz20}, where the goodness of fit was driven by the broader FeH feature in the $H$ band, thus allowing for a poorer fit in the $J$ band. For SDSS J1256$-$0224, we do not detect the $H$ band FeH feature as collisionally-induced H$_2$ absorption shapes the longer-wavelength end of the $H$ band, however, this too is likely driven by a trade off for better fitting the other parts of the spectrum.  

As seen for the d/sdL7 SDSS J1416$+$1348A in \cite{Gonz20} and for the unusually red L4.5 2MASSW J2224438-015852 (hereafter J2224$-$0158) in \cite{Burn21}, we find that the uniform-with-altitude gas abundance method is preferred over thermochemical equilibrium. This method appears to be preferred for sources that are outliers, those which are unusual in some way from standard field brown dwarfs, as it allows for more flexibility to fit a wide range of spectral features. Additionally, the uniform-with-altitude method provides flexibility to attempt to better fit spectral features that are ongoing challenges for the thermochemical equilibrium method such as the treatment of pressure broadening for the alkalies \citep{nallard16,nallard19}.

Figure \ref{fig:1256_spectrum_vmr}(b) shows retrieved abundances compared to the expected abundances from a [Fe/H]$_{ion}=-1.5$, C/O$=0.25x$ Solar thermochemical equilibrium model from the grid introduced in Section~\ref{sec:gas_abundances}. Here we see that our retrieved H$_2$O abundance matches what is expected from the grid models. For FeH and CrH, it is difficult to compare our abundances to the chemical models as these abundances vary greatly in the photosphere and are not close to uniform-with-altitude. The photosphere is shown as a gray strip to guide the reader on reasonable abundance ranges expected for the metal-hydrides. We find that the retrieved FeH abundance falls within a large range of expected photospheric abundances, while for CrH our 1$\sigma$ confidence interval just agrees with the highest possible abundance expected.

When we compare our retrieved gas abundances to those from a Solar C/O ratio, [Fe/H]$_{ion}=-1.5$ grid model, we found no major difference in the model values expected for FeH and CrH, however, for H$_2$O we retrieve an abundance slightly greater than expected by the chemical equilibrium grid models. As SDSS J1256$-$0224 is an old low-metallicity source we expect there to have been a lower abundance of metals available during its formation, which is apparent by our inability to retrieve any carbon-bearing species. Therefore the water abundance agreeing with the lower C/O ratio model is expected. Because of this we tested the impact of using a C/O ratio that is $0.25x$ Solar for the ions and found that it was indistinguishable from the winning model. Interestingly, when testing the thermochemical gas abundance method, we retrieved a C/O ratio near $0.25x$ Solar, however this model pushes the metallicity to the low end of the grid and retrieves a very low gravity which is not expected for an old source.

\section{Discussion}\label{sec:Discussion}

\subsection{Fundamental Parameter Comparison  \label{sec:FundParmComp1256}} 
\begin{figure*}
\gridline{\fig{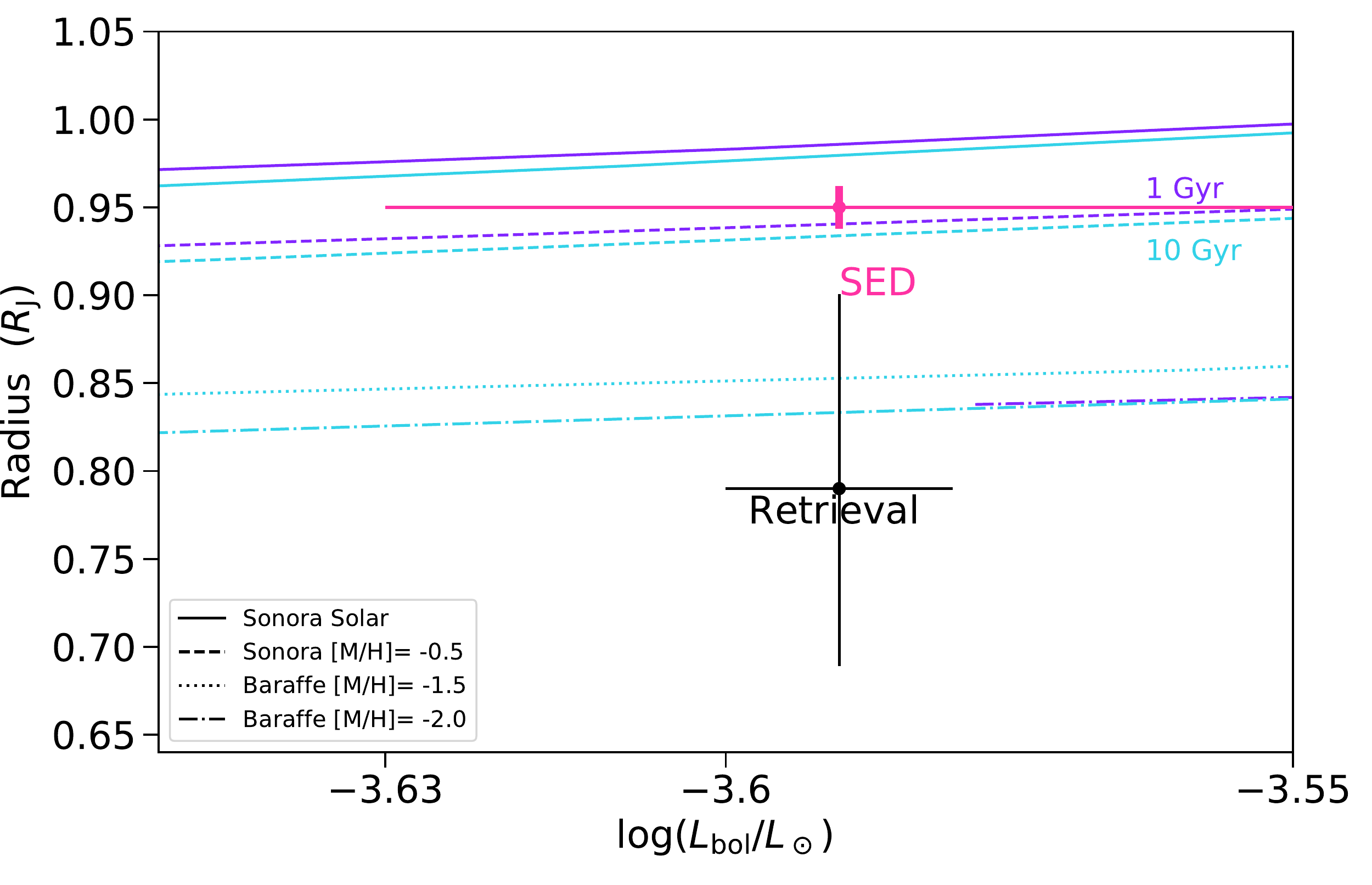}{0.5\textwidth}{\large(a)}
          \fig{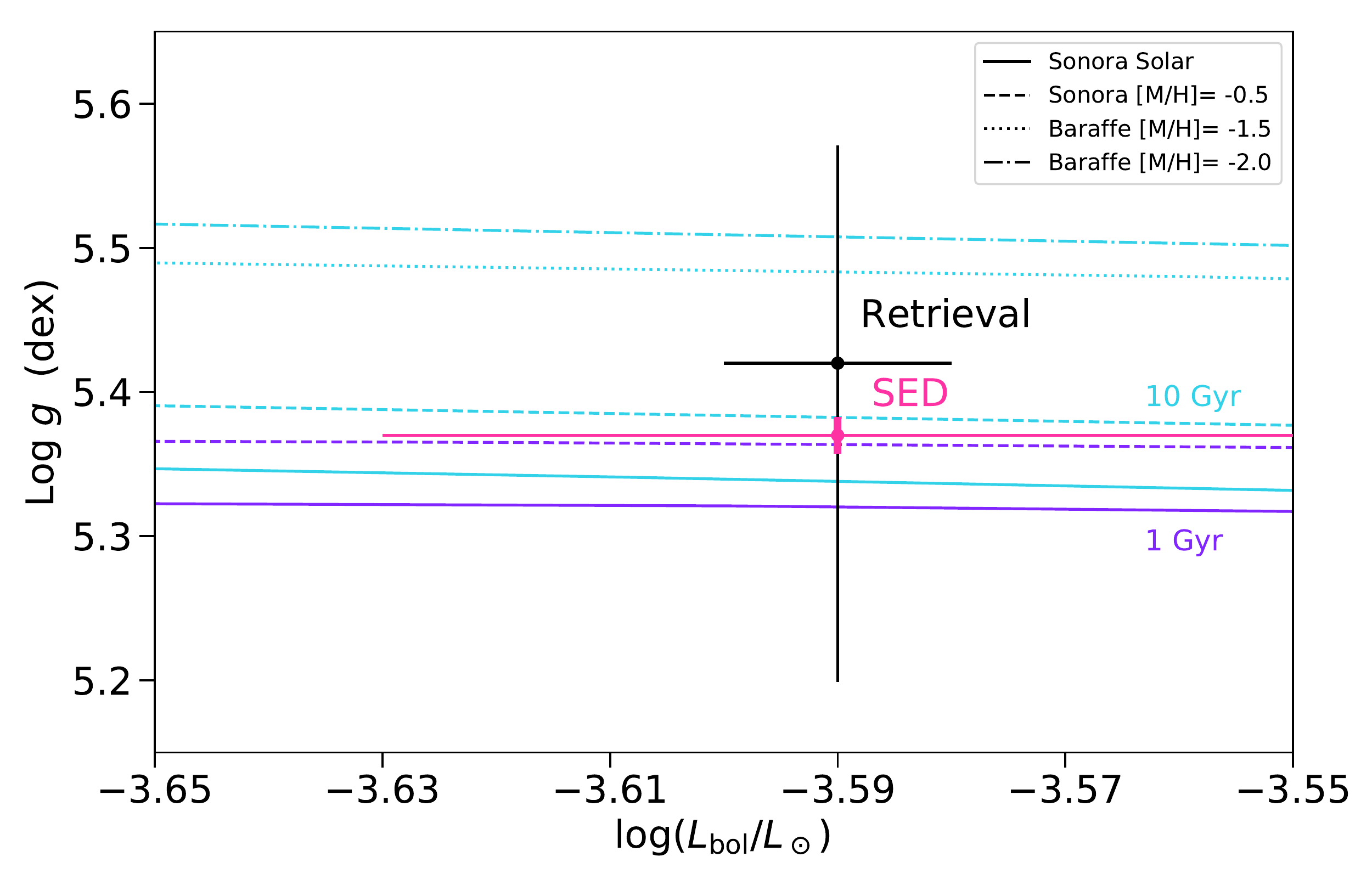}{0.5\textwidth}{\large(b)}}
 \gridline{\fig{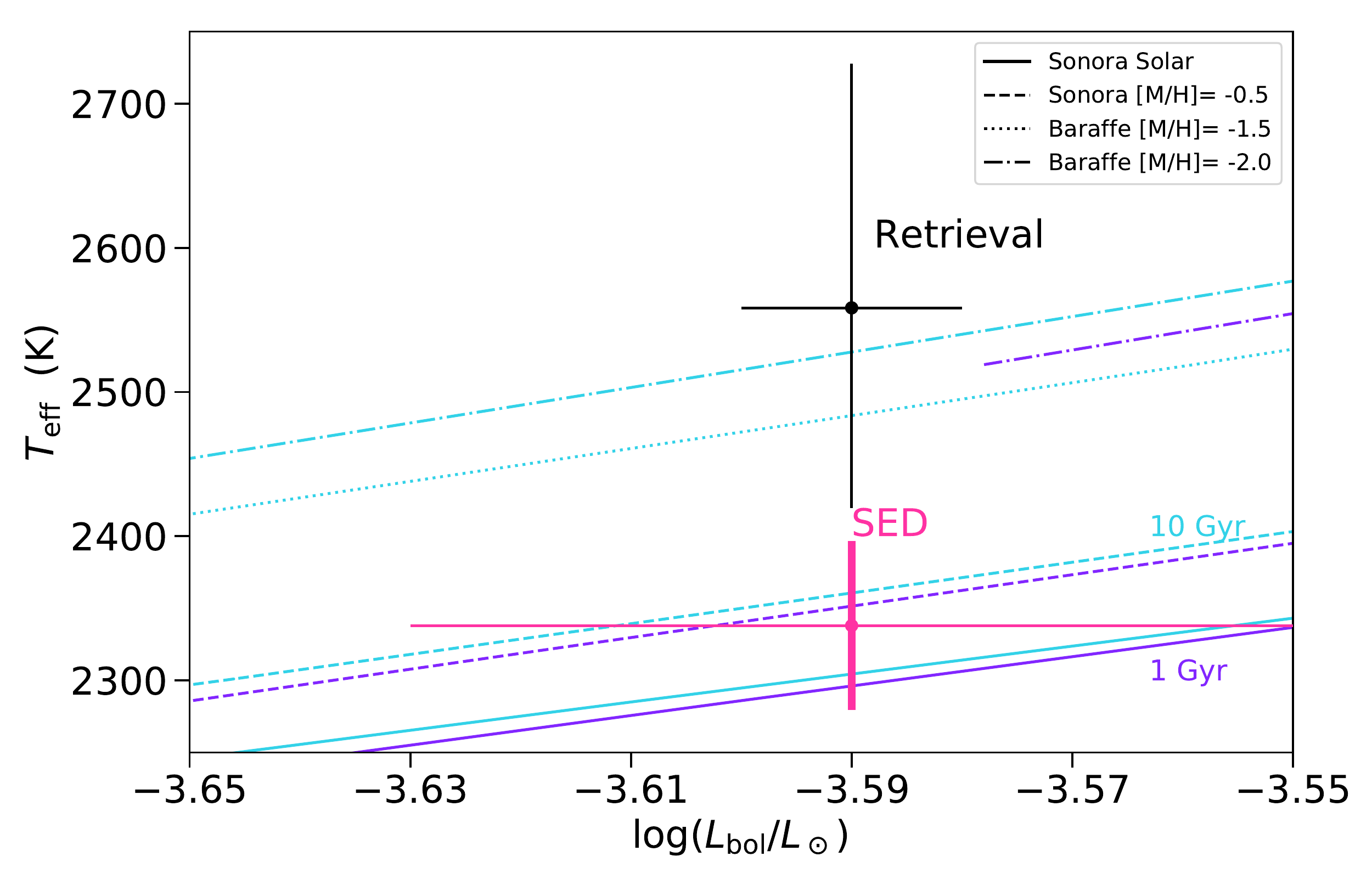}{0.5\textwidth}{\large(c)}
           \fig{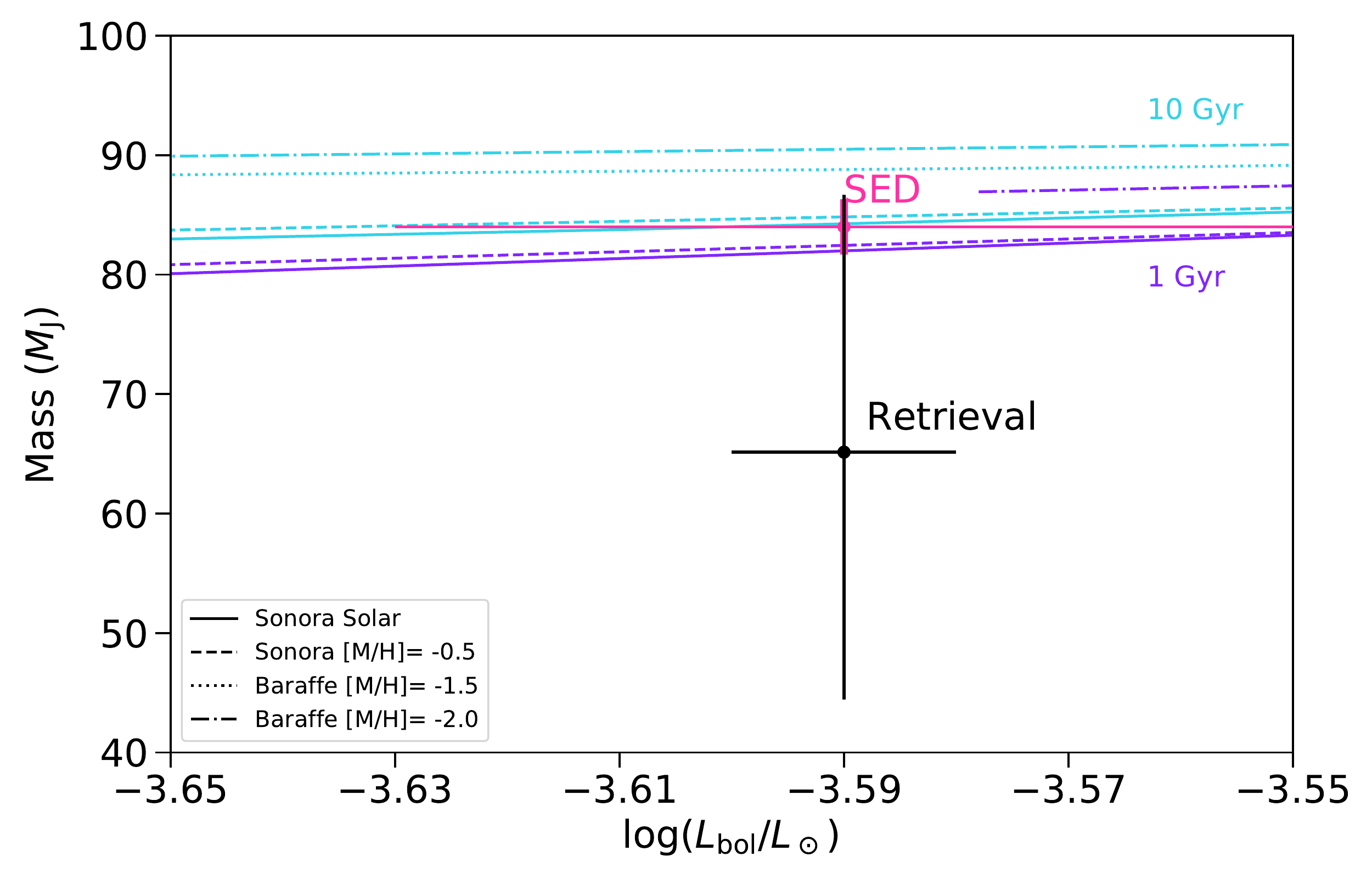}{0.5\textwidth}{\large(d)}} 
\caption{Comparison of retrieved bolometric luminosity, radius, surface gravity, \Teff, and mass to the Sonora Bobcat evolutionary Solar and low metallicity models, as well as the Baraffe low-metallicity models. {[M/H]}$=0.0$ are displayed as solid lines, {[M/H]}$=-0.5$ as dashed lines, {[M/H]}$=-1.5$ as dotted lines, and {[M/H]}$=-2.0$ as dashed-dot lines with the $1$~Gyr and $10$~Gyr curves in blue and purple respectively. For the SED values, the thick pink lines indicate a range as they use the radius range in the determination of the value, while thin lines are the uncertainties. (a) Radius versus \Lbol. (b) Log\,$g$ versus \Lbol. Log\,$g$ values come from \cite{Bara97} and are only available for an age of 10 Gyr. (c) \Teff versus \Lbol. (d) Mass versus \Lbol.}
\label{fig:Comparison_evo_models}
\vspace{0.5cm} 
\end{figure*}

Figures~\ref{fig:Comparison_evo_models}(a)--(d) compare our retrieval- and SED-based fundamental parameters of SDSS J1256$-$0224 to the Sonora Bobcat cloud-free evolutionary model grids for [M/H]$=0.0$ and [M/H]$=-0.5$ and the Baraffe evolutionary models \citep{Chab97, Bara97} for [M/H]$=-1.5$ and [M/H]$=-2.0$. The SED-based values for \Teff, mass, and radius are drawn from different evolutionary models (\citealt{Saum08} [M/H]=-0.3 grid, see \citealt{Gonz18} for additional details) and therefore are only shown as comparison to our retrieval-inferred values. We find that the SED- and retrieval-based log\,$g$, \Teff, and mass agree within 1$\sigma$ while the radius differs by 1.5$\sigma$. The retrieval-inferred radius, and \Teff agree with the Baraffe [M/H]$=-1.5$ and [M/H]$=-2.0$ models, while the log\,$g$ is consistent with all of the Sonora and Baraffe models. The mass is only consistent with the Sonora and Baraffe Solar and [M/H]$=-0.5$ models, however, when using the complete gas model the mass is consistent with all models. 

Unlike previous L dwarfs retrieved with \textit{Brewster}, SDSS J1256$-$0224 has a radius consistent with the evolutionary model radius, however, the median value is slightly smaller than expected from the Baraffe models. The retrieval-inferred radius match to the evolutionary models is likely due to the cloud-free nature of SDSS J1256$-$0224. The majority of T7 and T8 dwarfs retrieved in the literature have radii in agreement with evolutionary models. Using \textit{Brewster}, \cite{Gonz20} found the radius of SDSS J1416$+$1348B agreed with the low-metallicity Sonora models and the SED-based radius. Using CHIMERA, \cite{Line17} found 9 out of the 11 T7 and T8 dwarfs they retrieved agreed with the SED-based radii from \cite{Fili15}. Additionally, \cite{Kess18} found radii of M subdwarfs (sdM6 and earlier) were in agreement with the Baraffe model radii. They used a method similar to that of \cite{Fili15} to determine \Lbol, fit BT-Settl CFIST spectra to their observed data to get \Teff values, and then determined radii using the Stefan-Botlzmann law. With the sdL3.5 SDSS J1256$-$0224, this is the first sign that this agreement could follow for the cooler mid-L subdwarfs.

\subsection{How can retrieved abundances help inform the origins of a target?}
Understanding the chemical makeup of a target is critical when trying to decipher it's origin. For higher mass stars in the stellar halo, \cite{Helm18} found by cross-matching \textit{Gaia} DR2 and APOGEE, that two distinct chemical abundance sequences are visible on a color-magnitude diagram, which trace the Gaia-Enceladus merger. Particularly low [Fe/H] and high [$\alpha$/Fe] values as well as noticeably different chemical abundances than the thick disk or halo stars formed in-situ were found for the Gaia-Enceladus stars \citep{Helm18, Hayw18}. \cite{Hayw18} also noted that the low- and high-[$\alpha$/Fe] sequences of halo stars observed by \cite{Niss10} correspond to the blue and red sequences of the main-sequence in the Hertzsprung-Russel (HR) diagram for kinematically identified halo stars. Later, \cite{Di_Ma19} found that the blue sequence (low-[$\alpha$/Fe) likely originated from the accretion of stars in the Gaia-Enceladus merger and the red sequence (high-[$\alpha$/Fe]) corresponds to in-situ Milky Way Stars. 

For substellar subdwarfs, retrievals provide the only way for us to determine their chemical abundances. Including optical spectra ($\sim0.5-1.0\,\mu$m) into retrievals of subdwarfs, would provide additional metal-hydride features and alkali lines that could help to better constrain their retrieved gas abundances. However, optical data has been historically excluded in retrievals due to the strong impact pressure broadened wings of the $0.77\mu$m \ion{K}{1} doublet has on the pseudo-continuum. There have been recent updates to the treatment of alkali pressure broadening \citep{nallard16,nallard19} and by including these updates Gonzales et al. (in prep) are working on exploring the effect including optical data will have on the retrieved gas abundances. Therefore, with a large enough sample of substellar subdwarfs with well-constrained retrieved chemical abundances and kinematics, these low- and high-[$\alpha$/Fe] sequences could potentially be extended down to include the lowest mass objects enabling us to determine if any substellar subdwarfs are remnants of the Gaia-Enceladus merger.

\subsection{Retrieval Insights from the ``Rejected'' Models}  
The ``rejected'' models, those with a $\Delta$BIC$>45$, can provide a great deal of insight on how the retrieval model compensates to fit the data with a worse $\Delta$BIC model. Of particular interest is the behavior of the indistinguishable full gas set models, the cloud and its opacity in the cloudy models, and the behavior of the chemical equilibrium model. Additionally, the uniform-with-altitude models can inform us how much of an effect the choice of a poorer $\Delta$BIC cloud model can have on the retrieved gas abundances.

\subsubsection{Insights From Indistinguishable Full Gas Set Models  \label{sec:indistinguishable_model_comp}} 
From testing the full gas set models, we found that the cloudless ion metallicity [Fe/H]$_{ion}=-2.5,-2.0$ and $-1.5$ (drawn for a Solar C/O ratio grid) models as well as the cloudless [Fe/H]$_{ion}=-1.5$ (drawn from a C/O$=0.25x$ Solar grid) model were indistinguishable from one another. After running these models with the reduced gas set, the cloud-free [Fe/H]$_{ion}=-2.5$ model was no longer indistinguishable. Here we will discuss the behavior of the full gas set indistinguishable models as the corresponding reduced gas set models behave in a similar way.

\begin{figure*}
\gridline{\fig{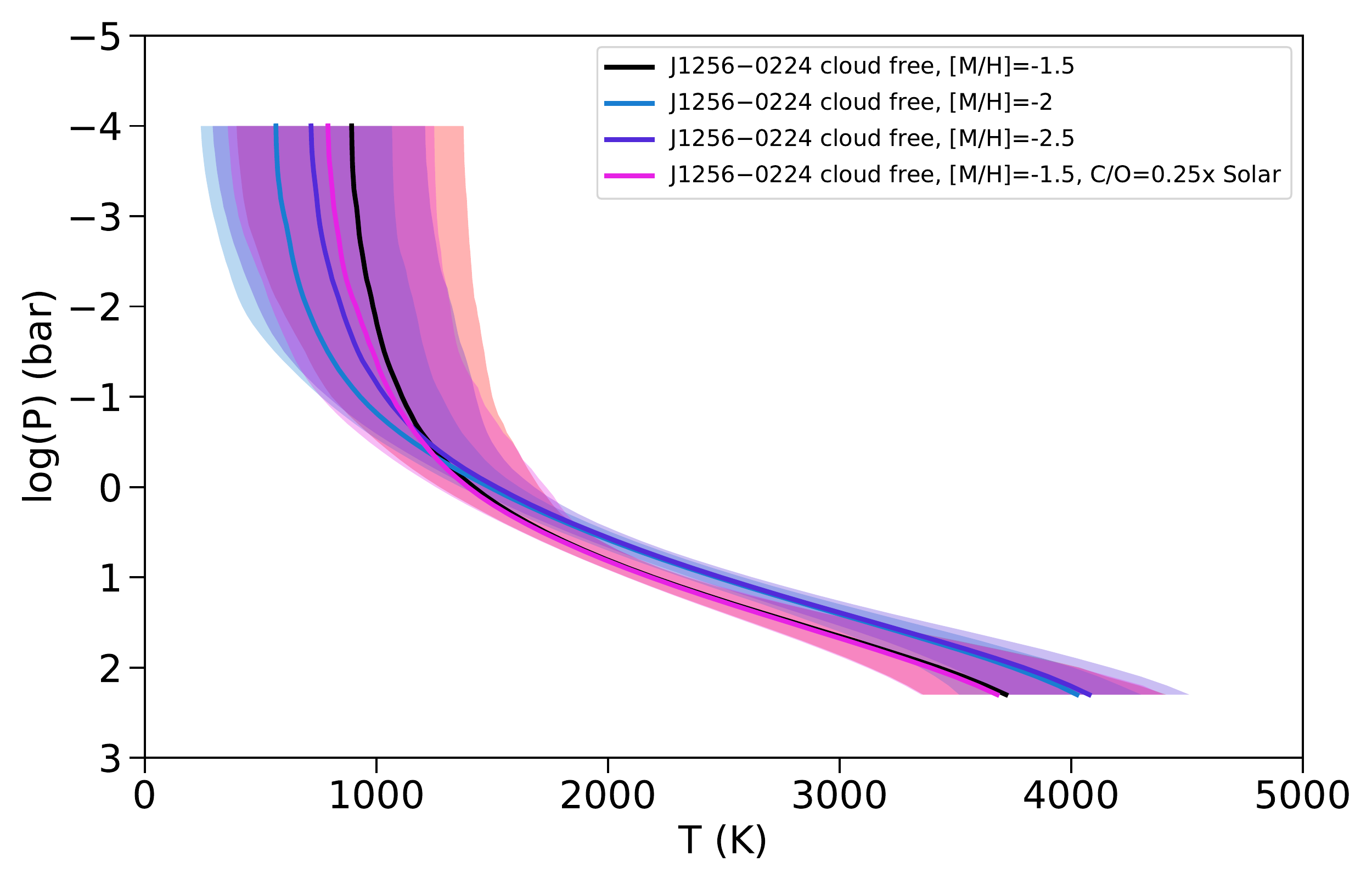}{0.5\textwidth}{\large(a)} 
          \fig{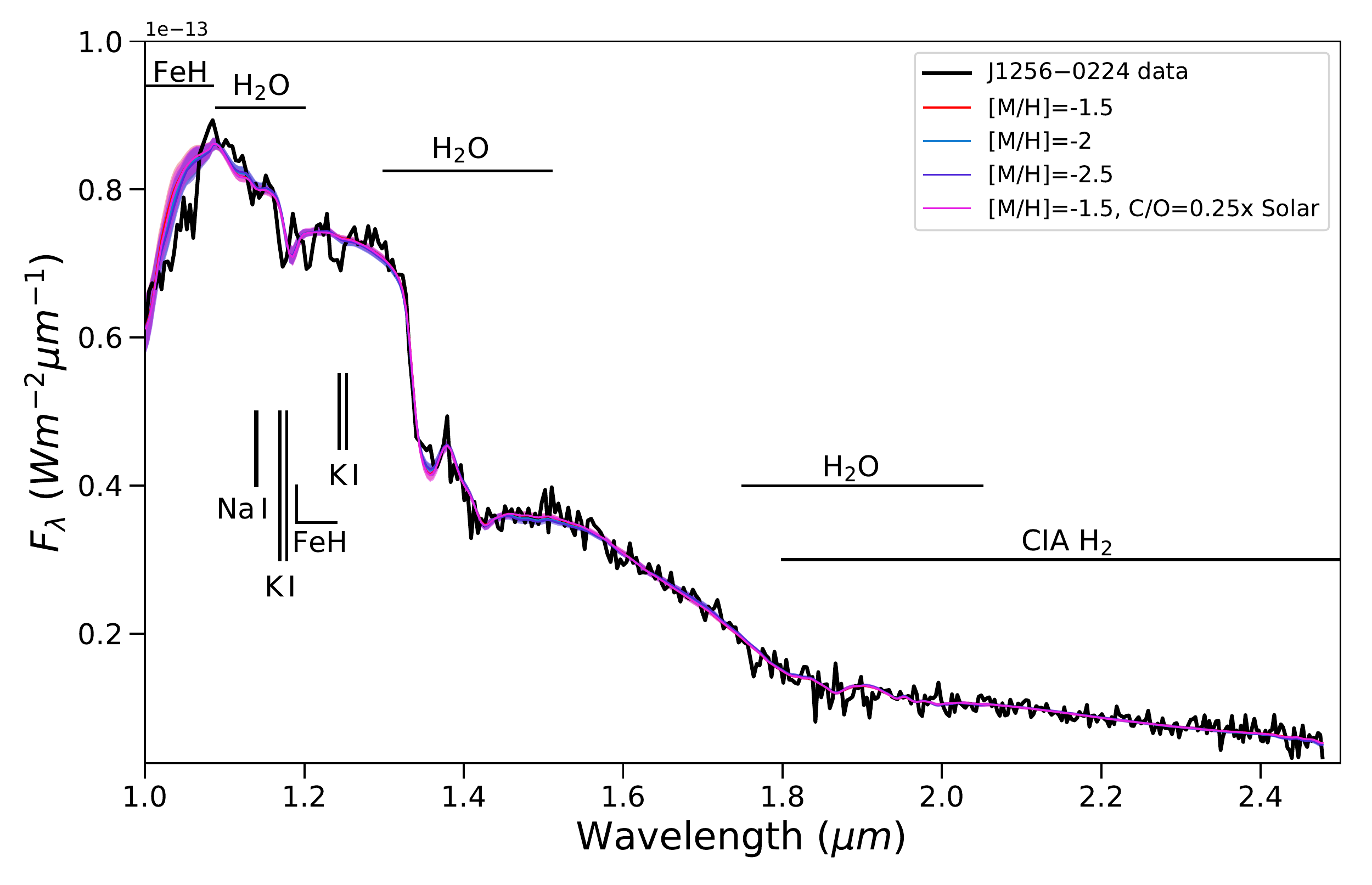}{0.5\textwidth}{\large(b)}} 
\caption{(a) Comparison of the P-T profiles for the three indistinguishable models for SDSS J1256$-$0224. Top model ([M/H]=-1.5) in red, ([M/H]=-1.5) in blue, and ([M/H]=-1.5) in purple. (b) Comparison on the resultant retrieval model spectra.}
\label{fig:1256_PTprofcomp}
\vspace{0.5cm} 
\end{figure*}

Figure \ref{fig:1256_PTprofcomp} compares the P-T profiles and the model spectra of the four indistinguishable full gas set models. In the photosphere (log$P\sim~0.5-2$~bars), the four profiles separate into two options-- the cooler profiles of the [Fe/H]$_{ion}=-1.5$ models with either Solar or 0.25$x$ Solar C/O and the hotter profiles for the [Fe/H]$_{ion}=-2$ and $-2.5$ ion metallicity models. At 1 bar, there is a $\sim300$~K temperature difference between two groupings of P-T profiles. In the upper atmosphere (lower pressures), the $1 \sigma$ contours for all four models overlap while the median profiles become cooler as you move to lower metallicities. Figure \ref{fig:1256_PTprofcomp}(b) shows the retrieved model spectra for each of the indistinguishable models. While they all look quite similar overall, slight differences occur in their fits to the the FeH feature in the $Y$ band as well as the H$_2$O features in the $J$ band.

\begin{deluxetable*}{l c c c c}
\tablecaption{Retrieved Gas Abundances and Derived Properties for SDSS J1256$-$0224 indistinguishable full gas set models\label{tab:Comp_corner_list}} 
\tablehead{\colhead{Parameter}& \multicolumn{4}{c}{Value for Ions for Continuum Opacities} \\
           \cmidrule(lr){2-5}\colhead{} & \multicolumn{2}{c}{[M/H]$=-1.5$} & \colhead{[M/H]$=-2$} & \colhead{[M/H]$=-2.5$}\\
           \cmidrule(lr){2-3}\colhead{} & \colhead{C/O=1.0} & \colhead{C/O=0.25} & \colhead{}} 
  \startdata
  \multicolumn{5}{c}{Retrieved} \\\hline
  H$_2$O & $-4.58\substack{+0.16 \\ -0.14}$ & $-4.57\substack{+0.18 \\ -0.16}$& $-4.71\substack{+0.20 \\ -0.19}$ & $-4.60\substack{+0.19 \\ -0.18}$\\ 
  CO & <$-5.63$ & <$-5.57$& <$-5.74$ & <$-5.67$ \\
  CO$_2$ & <$-4.79$ & <$-4.91$& <$-5.45$ & <$-5.49$\\
  CH$_4$ & <$-6.14$ & <$-5.92$& <$-6.39$ & <$-6.81$\\
  TiO & $<-9.75$ & $<-9.71$ & $<-9.88$ & $<-9.77$\\
  VO & $<-9.92$ & $<-9.98$& $<-9.87$ & $<-9.87$\\
  CrH & $-8.95\substack{+0.14 \\ -0.16}$ & $-8.95\substack{+0.16 \\ -0.19}$& $-9.08\substack{+0.18\\ -0.19}$ & $-9.03\substack{+0.20 \\ -0.19}$\\
  FeH & $-9.49\substack{+0.59 \\ -1.32}$ & $-9.43\substack{+0.56 \\ -0.97}$& $-9.41\substack{+0.45 \\ -0.73}$ & $-9.13\substack{+0.42 \\ -0.63}$\\
  Na+K & $<-8.64$ & $<-8.67$ & $<-8.65$ & $<-8.56$\\
  log\,$g$ (dex) & \phm{+}$5.44\substack{+0.19 \\ -0.20}$ & \phm{+}$5.47\substack{+0.17 \\ -0.24}$& \phm{+}$5.12\substack{+0.29 \\ -0.33}$ & \phm{+}$5.27\substack{+0.30 \\ -0.29}$\\ \hline 
  \multicolumn{5}{c}{Derived}  \\ \hline
  \Lbol & $-3.60 \pm 0.01$ & $-3.59 \pm 0.01$& $-3.59 \pm 0.01$ & $-3.60 \pm 0.01$ \\ 
  \Teff (K) & $2550.46\substack{+194.50 \\ -170.03}$ & $2538.43\substack{+233.35 \\ -152.20}$& $2648.53\substack{+171.58 \\ -177.44}$ & $2716.85\substack{+160.47 \\ -154.07}$ \\ 
  Radius ($R_\mathrm{Jup}$) & \phm{+}$0.79\substack{+0.13 \\ -0.12}$ & \phm{+}$0.80\substack{+0.13 \\ -0.15}$& \phm{+}$0.74\substack{+0.12 \\ -0.09}$ & \phm{+}$0.70\substack{+0.10 \\ -0.09}$ \\ 
  Mass ($M_\mathrm{Jup}$) & \phm{+}$72.11\substack{+21.75 \\ -24.24}$ & \phm{+}$74.09\substack{+20.23 \\ -24.40}$& \phm{+}$28.61\substack{+21.36 \\ -11.56}$ & \phm{+}$37.36\substack{+29.07 \\ -16.60}$\\
  {C/O} & $\cdots$ & $\cdots$ & $\cdots$ & $\cdots$ \\ 
  {[M/H]}\tablenotemark{a} & $-1.53\substack{+0.16 \\ -0.14}$ & $-1.51\substack{+0.18 \\ -0.16}$ & $-1.65\substack{+0.20 \\ -0.19}$ & $-1.55\substack{+0.19 \\ -0.18}$\\
  \enddata
  \tablenotetext{a}{Atmospheric value.}
  \tablecomments{Molecular abundances are fractions listed as log values. For unconstrained gases, 1$\sigma$ confidence is used to determine upper limit. C/O$=1.0$ is Solar abundance and C/O$=0.25$ is one quarter Solar abundance.}
\end{deluxetable*}

Table \ref{tab:Comp_corner_list} shows the retrieved abundances and extrapolated parameters for the four indistinguishable full gas set models. All of the retrieved and derived quantities are consistent between the models within the $1\sigma$ intervals. For the \Teff and radius, we find that as the ion metallicity decreases the median \Teff increases while the radius becomes smaller. Interestingly, moving from the [Fe/H]$_{ion}=-1.5$ model to the [Fe/H]$_{ion}=-2$ model, the median mass drops drastically and is smaller than expected by evolutionary models. This set of indistinguishable models tells us that while ion metallicity plays an important role in subdwarf retrievals, using an ion metallicity lower than the atmospheric metallicity does not have a major impact on the retrieved parameters but will strongly affect the derived mass. 

It is interesting that the cloud-free ion metallicity [Fe/H]$_{ion}=-2.5$ model is distinguishable when more gases are included in the retrieval model. The largest difference in the resultant spectrum between the full gas set and reduced gas set models is the seen in the $Y$ band FeH feature. In this region, the reduced gas set spectra lie further above the data than the full gas set spectra. Additionally of the gases in common, the FeH abundance median has the largest increase between the two gas sets. These FeH differences could play a role in producing distinguishable $\Delta$BIC vales for the reduced gas cloud-free ion metallicity [Fe/H]$_{ion}=-2.5$ model.

\subsubsection{Cloud Behavior}
For the cloudy retrievals, we only ran models using the full gas set indistinguishable models with ion metallicities of [Fe/H]$_{ion}=-1.5, -2, \mathrm{and} -2.5$. Regardless of the ion metallicity, the deck cloud models always pushed the cloud top to the bottom of the atmosphere and were unable to constrain the vertical extent of the cloud (where the opacity drops to $\tau=0.5$). With the power-law deck cloud, the power $\alpha$ was consistent with zero. For the grey deck cloud, the $\tau=1$ cloud opacity was always below the gas opacity near 100 bar across the entire spectral range, while the power-law deck cloud $\tau=1$ cloud opacity followed a similar behavior across the entire spectral range or until it became optically thin. For the slab cloud parameterization models, both the grey and power-law models were optically thin with median opacites ranging from $\sim0.2-0.4$ dex, with the median opacity increasing as the ion metallicity decreased. The location and thickness of the slab clouds were entirely unconstrained. For all of the cloud models, the retrieved spectrum looks, by eye, similar to the winning cloud-free spectrum as the clouds were pushed out of sight making them all effectively cloud-free.

\subsubsection{Thermochemical grid model}
As part of our model sample, we tested the thermochemical equilibrium gas abundance method in addition to the uniform-with-altitude method to determine which method was preferred. The thermochemical equilibrium method was only tested with cloud-free, solar ion metallicity model. In this model, the P-T profile was consistent with the Sonora grid profiles. We retrieve a metallicity at the bottom edge of our prior along with a subsolar C/O ratio and a log\,$g$ which would indicate a low surface gravity-- an unlikely result for an old source. The rejection of this model is influenced by its reduced flexibility in comparison to the uniform-with-altitude method, which is visible in the poor fit of the retrieved spectrum to the observed data in the $Y$, $J$, and $H$ bands.

\subsubsection{Impact on Retrieved Gas Abundances}
A remaining question for brown dwarf retrievals is ``Are we able to retrieve gas abundances matching with the winning model even if we have a poorer fitting model (lower $\Delta$BIC)?'' From examining all the models tested for SDSS J1256$-$0224, we found that for a majority of the models, the gas abundances are not affected. Table \ref{tab:Rejects} lists the abundances for the constrained gases and log\,$g$ for all of the rejected uniform-with-altitude models we tested in order of increasing $\Delta$BIC. As seen in Table~\ref{tab:Rejects}, the only model with gas abundances and a log\,$g$ that are inconsistent with the winning model is the cloud-free Solar ion metallicity model, the poorest model with a $\Delta$BIC$=129.2$. Interestingly, this model is the only one that constrains abundances for CO$_2$ and CH$_4$ requiring them to be the two most abundant gases respectively. So in the case of SDSS J1256$-$0224, you can have a poorer fitting model as long as the ion metallicity is subsolar and get gas abundances consistent with the indistinguishable ``winning'' models.

\begin{deluxetable*}{l c c c c}
\tablecaption{Rejected Models Retrieved Gas Abundances and log\,$g$ Comparison\label{tab:Rejects}} 
\tablehead{\colhead{Model\tablenotemark{a}}& \colhead{H$_2$O} & \colhead{CrH} & \colhead{FeH} &  \colhead{log\,$g$}}
  \startdata
  Cloud Free\tablenotemark{b}, [M/H]$=-2.5$ & $-4.60\substack{+0.19 \\ -0.18}$  & $-9.03\substack{+0.20 \\ -0.19}$  & $-9.13\substack{+0.42 \\ -0.63}$  & $5.27\substack{+0.30 \\ -0.29}$  \\
  Cloud Free,[M/H]$=-1.5$ & $-4.58\substack{+0.16 \\ -0.14}$  & $-8.95\substack{+0.14 \\ -0.16}$  & $-9.49\substack{+0.59 \\ -1.32}$  & $5.44\substack{+0.19 \\ -0.20}$  \\
  Cloud Free, uniform,[M/H]$=-1.5$, C/O$=0.25x$ & $-4.57\substack{+0.18 \\ -0.16}$  & $-8.95\substack{+0.16 \\ -0.19}$  & $-9.43\substack{+0.56 \\ -0.97}$  & $5.47\substack{+0.17 \\ -0.24}$  \\
  Cloud Free, [M/H]$=-2.5$ & $-4.60\substack{+0.19 \\ -0.18}$ & $-9.03\substack{+0.20 \\ -0.19}$ & $-9.13\substack{+0.42 \\ -0.63}$ & $5.27\substack{+0.30 \\ -0.29}$\\
  Cloud Free, [M/H]$=-2.0$  & $-4.71\substack{+0.20 \\ -0.19}$ & $-9.08\substack{+0.18\\ -0.19}$ & $-9.41\substack{+0.45 \\ -0.73}$  & $5.12\substack{+0.29 \\ -0.33}$ \\ 
  Cloud Free, [M/H]$=-1.0$  & $-4.56\substack{+1.26 \\ -0.11}$  & $-8.90\substack{+0.75 \\ -0.13}$ & $-10.12\substack{+3.18 \\ -1.29}$ & $5.47\substack{+0.13 \\ -0.14}$ \\ 
  Grey Deck cloud, [M/H]$=-1.5$ & $-4.61\substack{+0.15 \\ -0.13}$ & $-8.97\substack{+0.83 \\ -0.16}$ & $-9.70\substack{+0.57 \\ -1.30}$ & $5.39\substack{+0.17 \\ -0.19}$ \\
  Grey Deck cloud, [M/H]$=-2.0$ & $-4.69\substack{+0.23 \\ -0.18}$ & $-9.10\substack{+0.19 \\ -0.20}$ & $-9.50\substack{+0.53 \\ -0.96}$ & $5.13\substack{+0.27 \\ -0.33}$ \\
  Grey Deck cloud, [M/H]$=-2.5$ & $-4.91\substack{+0.22 \\ -0.18}$ & $-9.38\substack{+0.24 \\ -0.23}$ & $-9.46\substack{+0.46 \\ -0.38}$ & $4.71\substack{+0.43 \\ -0.41}$ \\
  Grey Slab cloud, [M/H]$=-2.0$ & $-4.65\substack{+0.14 \\ -0.16}$ & $-9.04\substack{+0.14 \\ -0.19}$ & $-9.26\substack{+0.38 \\ -0.69}$ & $5.20\substack{+0.24 \\ -0.28}$ \\ 
  Power-law Deck cloud, [M/H]$=-1.5$ & $-4.67\pm 0.16$ & $-9.02\substack{+0.15 \\ -0.23}$ & $-9.66\substack{+0.52 \\ -1.31}$ & $5.22\substack{+0.25 \\ -0.30}$ \\
  Grey Slab cloud, [M/H]$=-1.5$ & $-4.65\substack{+0.12 \\ -0.11}$ & $-9.00\substack{+0.13 \\ -0.15}$ & $-9.79\substack{+0.57 \\ -1.25}$ & $5.32\substack{+0.17 \\ -0.21}$ \\
  Grey Slab cloud, [M/H]$=-2.5$ & $-4.86\substack{+0.33 \\ -0.18}$ & $-9.36\substack{+0.29 \\ -0.27}$ & $-9.17\substack{+0.58 \\ -1.51}$ & $4.58\substack{+0.46 \\ -0.34}$ \\
  Power-law Deck cloud, [M/H]$=-2.0$ & $-4.74\substack{+0.19 \\ -0.20}$ & $-9.14\substack{+0.23 \\ -0.26}$ & $-9.60\substack{+0.50 \\ -0.95}$ & $5.10\substack{+0.32 \\ -0.37}$ \\
  Power-law Deck cloud, [M/H]$=-2.5$ & $-4.90\substack{+0.22 \\ -0.19}$ & $-9.34\substack{+0.22 \\ -0.27}$ & $-9.57\substack{+0.43 \\ -0.59}$ & $4.71\substack{+0.47 \\ -0.41}$ \\
  Power-law Slab cloud, [M/H]$=-2.0$ & $-4.65\substack{+0.16 \\ -0.19}$ & $-9.07\substack{+0.17 \\ -0.23}$ & $-9.31\substack{+0.41 \\ -0.83}$ & $5.20\substack{+0.27 \\ -0.38}$ \\
  Power-law Slab cloud, [M/H]$=-2.5$ & $-4.75\substack{+0.19 \\ -0.21}$ & $-9.23\substack{+0.23 \\ -0.25}$ & $-9.28\substack{+0.41 \\ -0.50}$ & $5.00\substack{+0.35 \\ -0.59}$ \\  
  Power-law Slab cloud,  [M/H]$=-1.5$ & $-4.67\substack{+0.14 \\ -0.13}$ & $-9.04\substack{+0.16 \\ -0.20}$ & $-9.47\substack{+0.45 \\ -1.05}$ & $5.24\substack{+0.24 \\ -0.33}$ \\
  Cloud Free, [M/H]$=0.0$ & $-2.67\substack{+0.10 \\ -0.11}$ & $-7.40\substack{+0.21 \\ -0.28}$ & $-6.24\pm 0.13$ & $5.77\substack{+0.16 \\ -0.29}$ \\
  \enddata
  \tablenotetext{a}{Gas abundance method is uniform-with-altitude.}
  \tablenotetext{b}{Reduced gas set.}
  \tablecomments{Models listed in order of increasing $\Delta$BIC.}
\end{deluxetable*}

\subsection{Impact of Signal-to-Noise on the retrieved gas abundances}

An important aspect to consider is the signal-to-noise (SNR) of our spectrum and how that may impact what gases we are able to constrain. Previously L dwarfs retrieved with \textit{Brewster} have had a higher SNR in the $Y$ and $J$ bands, where we see the largest discrepancies in our retrieved spectrum to the observed data, than that of SDSS J1256$-$0224. For example in the peak of the $J$ band, the spectrum used in \cite{Gonz20} for SDSS J1416$+$1348A has an SNR$=437$ where SDSS J1256$-$0224 has an SNR$=84$. From examining another L subdwarf of similar metallicity and spectral type with an SNR=$492$ in the peak of the $J$ band, Gonzales et. al (in prep) have only been able to retrieve one additional gas. Therefore, the small amount of constrained gases is likely due to the low-metallicity of the source and not strongly dependent on the SNR of the spectrum.

\section{Conclusions}
In this work we present the first retrieval of a substellar subdwarf, SDSS J1256$-$0224, and provide an updated distance-calibrated SED for it. We find that SDSS J1256$-$0224 is best fit by a cloud-free model with a subsolar ion metallicity of [Fe/H]$_{ion}=-1.5$, however, this model is indistinguishable from cloud-free models with an ion metallicity of [Fe/H]$_{ion}=-2$ and a cloud-free [Fe/H]$_{ion}=-1.5$ drawn from a C/O$=0.25x$ Solar model. From our winning model, we are able to constrain gas abundances for H$_2$O, FeH, and CrH but are unable to constrain any of the carbon-bearing species. We find our extrapolated fundamental parameters from our retrieval agree with the low-metallicity \cite{Bara97} evolutionary models.  

Examinations of the ``rejected'' models, shows that gas abundances which match the ``winning'' model can be derived with a poorer fitting model, as long as the ion metallicity is subsolar. In the cloudy models, the retrieval will try to remove the cloud's effect by pushing the cloud to the bottom of the atmosphere or making it optically thin so it has very little to no contribution to the observed spectrum in the photosphere.

The results of this work help to confirm the cloud-free nature of subdwarfs and are part of a larger sample addressing the nature of how subdwarf P-T profiles compare to similar spectral type or \Teff sources (Gonzales et al. in prep). With additional retrievals of substellar subdwarfs, we can better address these questions as well as collect key abundance information that can aid in understanding the origins of these sources. 

\clearpage
\newpage

\acknowledgments
This research was supported by the NSF under Grant No. AST-1614527, grant No. AST-1313278, and grant No. AST-1909776. This research was made possible thanks to the Royal Society International Exchange grant No. IES{\textbackslash}R3\textbackslash170266. E.G. and J.F. acknowledge support from the Heising-Simons Foundation. BB acknowledges financial support from the European Commission in the form of a Marie Curie International Outgoing Fellowship (PIOF-GA-2013-629435). This work has made use of the University of Hertfordshire's high-performance computing facility. This publication makes use of data products from the Two Micron All Sky Survey, which is a joint project of the University of Massachusetts and the Infrared Processing and Analysis Center/California Institute of Technology, funded by the National Aeronautics and Space Administration and the National Science Foundation. This publication makes use of data products from the Wide-field Infrared Survey Explorer, which is a joint project of the University of California, Los Angeles, and the Jet Propulsion Laboratory/California Institute of Technology, funded by the National Aeronautics and Space Administration. This work has made use of data from the European Space Agency (ESA) mission {\it Gaia} (\url{https://www.cosmos.esa.int/gaia}), processed by the {\it Gaia} Data Processing and Analysis Consortium (DPAC, \url{https://www.cosmos.esa.int/web/gaia/dpac/consortium}). Funding for the DPAC has been provided by national institutions, in particular the institutions participating in the {\it Gaia} Multilateral Agreement.

\software{astropy \citep{Astropy},  
          SEDkit (\url{https://github.com/hover2pi/SEDkit}, \textit{Eileen Branch}), 
          \textit{Brewster} \citep{Burn17},
          EMCEE \citep{emcee},
          Corner \citep{Corner}
          }

\clearpage
\newpage

\bibliographystyle{yahapj}
\bibliography{references}

\end{document}